\documentclass[a4paper,fleqn,usenatbib]{mnras}

\usepackage{newtxtext,newtxmath}

\usepackage[T1]{fontenc}
\usepackage{ae,aecompl}

\usepackage{graphicx}	
\usepackage{amsmath}	
\usepackage{amssymb}	

\usepackage{amssymb}
\usepackage{lipsum}
\usepackage{afterpage}
\usepackage{threeparttable}
\usepackage{placeins}
\usepackage{multirow,makecell,booktabs}
\usepackage{subfigure}
\usepackage[nameinlink]{cleveref}
\Crefname{figure}{Fig.}{Figs.}
\usepackage{enumitem}

\newcommand{\hi}{\mbox{H{\sc i}}}

\newcommand{\kms}{km s$^{-1}$}

\newcommand{\ml}{$M_\odot/L_\odot$}

\newcommand{\Mo}{\rm M_{\odot}}
\newcommand{\Ms}{{\rm M_\star}}

\newcommand{\dg}{^{\circ}}
\newcommand{\mjybeam}{\rm mJy\,beam^{-1}}

\newcommand{\cm}{\rm cm^{-2}}

\newcommand{\dhr}{\overset{{\rm h}}{~}}

\newcommand{\dm}{\overset{{\rm m}}{~}}
\newcommand{\ds}{\overset{{\rm s}}{.}}

\newcommand{\dmin}{\overset{\prime}{.}}

\newcommand{\marc}{mag\,arcsec$^{-2}$}

\newcommand{\e}[1]{\times 10^{#1}}
\newcommand{\HI}{\textsc{Hi}}

\newcommand{\gipsy}{\textsc{Gipsy}}
\newcommand{\barolo}{\textsc{3d Barolo}}
\newcommand{\rotcur}{\textsc{Rotcur}}
\newcommand{\rotmas}{\textsc{Rotmas}}

\newcolumntype{H}{>{\setbox0=\hbox\bgroup}c<{\egroup}@{}} 

\title[\hi\ distribution \& kinematics of IC 2574 \& HIJASS J1021+68]{A $5\dg\times5\dg$ deep \hi\ survey of the M81 group: II. \hi\ distribution and kinematics of IC 2574 and HIJASS J1021+68}

\author[Sorgho et al.]{
A. Sorgho$^{1}$\thanks{sorgho@ast.uct.ac.za},
L. Chemin$^{2}$,
Z. S. Kam$^{3}$,
T. Foster$^{4,5}$,
C. Carignan$^{1,3}$
\\
\\
$^{1}$ Department of Astronomy, University of Cape Town, Private Bag X3, Rondebosch 7701, South Africa\\
$^{2}$ Centro de Astronom\'{i}a (CITEVA), Universidad de Antofagasta, Avenida Angamos 601, Antofagasta, Chile\\
$^{3}$ Laboratoire de Physique et de Chimie de l'Environnement, Observatoire d'Astrophysique de l'Universit\'{e} Joseph Ki-Zerbo (ODAUO),\\ \,\,\, 03 BP 7021, Ouaga 03, Burkina Faso\\
$^{4}$ Dominion Radio Astrophysical Observatory, P.O. Box 248, Penticton, British Columbia, V2A 6J9, Canada\\
$^{5}$ Department of Physics \& Astronomy, Brandon University, Brandon, MB R7A 6A9, Canada
}

\date{Accepted 2020 February 9.}

\pubyear{2020}

\begin{document}
\label{firstpage}
\pagerange{\pageref{firstpage}--\pageref{lastpage}}
\maketitle

\begin{abstract}
We analyse the eastern region of a $5\dg\times5\dg$ deep \hi\ survey of the M81 group containing the dwarf galaxy IC 2574 and the \hi\ complex HIJASS J1021+68, located between the dwarf and the M81 system. The data show that IC 2574 has an extended \hi\ envelope that connects to HIJASS J1021+68 in the form of a collection of small clouds, but no evident connection has been found between IC 2574 and the central members of the M81 group. We argue, based on the morphology of the clouds forming HIJASS J1021+68 and its velocity distribution, that the complex is not a dark galaxy as previously suggested, but is instead a complex of clouds either stripped from, or falling onto the primordial \hi\ envelope of IC 2574.
We also use the deep \hi\ observations to map the extended \hi\ envelope around IC 2574 and, using a 3D tilted-ring model, we derive the rotation curve of the galaxy to a larger extent than has been done before. Combining the obtained rotation curve to higher resolution curves from the literature, we constrain the galaxy's dark matter halo parameters.
\end{abstract}



\section{Introduction}
The most widely accepted models of galaxy formation, the standard $\Lambda$ cold dark matter ($\Lambda$CDM) and the hierarchical clustering model, predict that massive early type galaxies form through mergers of smaller, late type galaxies \citep[e.g.,][]{Peebles1965,Press1974,Searle1978,Blumenthal1984}. More generally, the hierarchical clustering scenario predicts that the smallest dark matter (DM) halos form first, and later merge to form larger structures, ranging from spiral galaxies to galaxy clusters (bottom-up scenario). Although the predictions of the models are consistent with observations on scales of galaxy clusters, there tends to be a discrepancy on smaller scales. In fact, the models predict more abundant low-mass halos, mostly in the form of satellites around giant spirals, than are revealed by observations \citep[the so-called `missing satellites' problem,][]{Klypin1999,Moore1999}. Furthermore, it has been suggested that high velocity clouds (HVCs), or at least one of their components (the so-called compact HVCs or CHVCs), might be the gaseous counterparts of low-mass dark-matter satellites \citep{Braun1999,Blitz1999}. This implies that numerous low column density gas clouds have yet to be detected in the local environment.

Over the past few years, effort has gone into mapping the low column density \hi\ clouds both in the vicinity of nearby galaxies and in galaxy groups \citep[see e.g.,][]{Braun2003,Braun2004,Chynoweth2008,Westmeier2008,Westmeier2013,Pisano2014}, in the hope of revealing a more complete picture of the distribution of the faint gas in the local Universe. Although several observational campaigns were successful in revealing faint neutral gas clouds, it was often found that limitations of the currently available instruments (such as poor resolution) could lead to an inaccurate picture of the \hi\ distribution \citep[e.g. the M31--M33 \hi\ filament,][]{Wolfe2013}. With the advent of the Square Kilometre Array precursor telescopes such as MeerKAT \citep{Camilo2018} and ASKAP \citep{Johnston2007}, the quality of \hi\ mapping of large-scale structures is expected to improve, with upcoming observations combining high resolutions, high sensitivities and large fields of view. For example, MeerKAT's MHONGOOSE survey \citep{DeBlok2018} aims to investigate cold gas accretion in isolated galaxies by mapping their \hi\ to very low column densities.

In galaxy groups, intergalactic \hi\ clouds in the form of tidal streams and tails are usually produced by interaction between galaxies \citep[e.g.,][]{Yun1993,Walter2002a,Putman2003}. It is also believed that some HVCs are the remnants of galaxy formation that are currently being accreted onto galaxies \citep[see][]{Maller2004}, or a result of a galactic fountain \citep[see][for a review]{Wakker1997}. Their existence in the surroundings of galaxies therefore gives insights into the formation processes of large-scale structures.


Using the DRAO telescope \citep{landecker2019}, we have mapped, to a high column density sensitivity, a $5\dg\times5\dg$ region of the M81 group covering the M81 interacting system (M81, M82 and NGC 3077), NGC 2976 and IC 2574 \citep[][hereafter Paper I]{Sorgho2019a} during ${\sim}3000$ hours of observation. The observations allowed us to {\it i)} map the \hi\ arm connecting the M81 system and NGC 2976, {\it ii)} resolve the \hi\ clouds adjacent to the arm, and also {\it iii)} resolve, for the first time, the HIJASS J1021+68 cloud located west of IC 2574 (see Fig. 6 of Paper I). In this second paper of the series, we focus on the subset of the observations data covering IC 2574 and HIJASS J1021+68 -- both located in the M81 group and at a distance of ${\sim}4$ Mpc \citep{Karachentsev2002} -- and perform additional observations to increase the sensitivity of the data in the region. 
As a dwarf galaxy presenting a slowly rising rotation curve (as compared to massive spirals), IC 2574 is an interesting object to study in the context of understanding the dark matter (DM) distribution in galaxies. In fact, several observational studies have shown that late-type dwarf galaxies are usually best described by a cored profile \citep[e.g.,][]{Moore1994,Flores1994,DeBlok2001,Swaters2003,Oh2011}, that is, their DM halos are better characterised by an approximately constant density core. This is in contrast with the $\Lambda$CDM model, which predicts a steeply cusped DM density distribution \citep{Dubinski1991,Navarro1996,Navarro1997,Moore1999}. A review of this discrepancy between observations and the model, known as the ``cusp-core'' problem, is given in \citet{DeBlok2010}.

We present both the extended emission around IC 2574, and a kinematical analysis of its \hi\ gas. We take advantage of the high column density sensitivity of the observations to extend the rotation curves of the galaxy found in the literature. We also present and discuss the \hi\ distribution and kinematics of HIJASS J1021+68.
The paper is organised as follows: in \Cref{sec:obs} we summarise the observations and the data reduction method and present the global \hi\ profiles and moment maps of the galaxy in \Cref{sec:hi-properties}. We next derive its rotation curve and construct its mass models in \Cref{sec:kinematics}, and summarise the results in \Cref{sec:summary}.

\section{Observations and data reduction}\label{sec:obs}
The observations were conducted in 2016/2017 using the DRAO telescope, and are extensively described in Paper I. Twenty different fields covering the IC 2574 galaxy, M81 and its interacting counterparts M82 and NGC 3077, as well as NGC 2976 were observed for 144 hours each during the observational campaign, for a total of nearly 3000 hours in total. To further increase the sensitivity of the data in certain key areas such as the regions containing respectively the M81 triplet and IC 2574, 10 additional fields were retrieved from the DRAO archive and added to the observations, bringing the total number of fields to 30. These fields were included in the mosaic presented in Paper I. Moreover, 8 additional fields centred on the region around IC 2574 were recently observed and added to the initial data, and therefore the final dataset comprises 38 fields, for a total of 5472 hours of observation. The fields were calibrated, reduced and processed individually before being combined into a single mosaic of $1024\times1024\times256$ pixels in size, with spatial and spectral pixel widths of $20''$ and $3.3$ \kms, respectively.
Because of the limitation of the DRAO as an interferometer when it comes to accurately recover the fluxes of large-scale structures (in this case, structures larger than $56'$), we have combined the data with EBHIS \hi\ data \citep{Winkel2016} to correct for the instrumental effect of missing the shortest spacings. The EBHIS data were obtained with the 100m Effelsberg single-dish telescope, and were shown to present a good overlap with the DRAO data in the {\it uv} plane. The resulting data has a $1\sigma$ sensitivity of $2.7\,\mjybeam$ around IC 2574, for a spatial resolution of $61''\times59''$ and a spectral resolution of 5.2 \kms. In column density units, this is equivalent to $5.8\e{18}\,\cm$ over 16 \kms\ and at a spatial resolution of 1.2 kpc. To further increase the sensitivity of the observations, we smoothed the datacube down to a resolution of $108''\times108''$ (or 2.1 kpc at the distance of IC 2574).

The last step of the data processing consisted of subtracting the Milky Way \hi\ emission from the IC 2574 emission. Because of the low systemic velocity of the galaxy, the data was contaminated by foreground Galactic \hi\ emission. We adopted a technique similar to that described in \citet{Chemin2009}, which consisted of subtracting a model of the Galactic emission from the data. The technique was successful in mitigating most of the foreground \hi\ in the data.
A complete description of the observations, the calibration and post-calibration processing procedures can be found in Paper I.
The smoothed cube, which is used in this work, has a $1\sigma$ noise level of $2.4\e{18}\,\cm$ (around IC 2574) over 16 \kms\ and a spatial resolution of 2.1 kpc (at the distance of IC 2574).

\begin{table}
	\begin{tabular}{p{5cm} l}
	\hline
    \hline
    Parameter & Value \\
    \hline
    R.A (J2000)\dotfill & $10\dhr28\dm23\ds5$\\
    Dec. (J2000)\dotfill & $68^\circ24'43.7''$\\
	Morphological type $^{\rm a}$\dotfill & SABm\\
	Distance$^{\rm b} $\dotfill & 4.02 Mpc\\
	Optical 25th mag. diameters $^{\rm c}$\dotfill & $13\dmin2\times5\dmin4$\\
	Total B Magnitude $^{\rm c}$\dotfill & 10.84 mag\\
	Stellar mass $^{\rm d}$\dotfill & $0.4\e{9}\,\Mo$\\
	Inclination $^{\rm e}$\dotfill & $75\dg$\\
	Position angle $^{\rm f}$\dotfill & $50\dg$\\
    \hline
	\end{tabular}
	\caption{Optical parameters of IC 2574. References: $^{\rm a}$RC3 catalogue \citep{DeVaucouleurs1991}; $^{\rm b}$\citet{Karachentsev2002a}; $^{\rm c}$\citet{Karachentsev2004}; $^{\rm d}$Derived from a {\it WISE} photometry (Jarrett et al., {\it in prep.}); $^{\rm e}$\citet{Salo2015}; $^{\rm f}$\citet{Appleton1981}.}
\end{table}

\section{\hi\, properties}\label{sec:hi-properties}

\subsection{\hi\ emission in the M81 group}\label{sec:m81-emission}
\begin{figure*}
\centering
\makebox[\textwidth][c]{
\includegraphics[width=1.\textwidth]{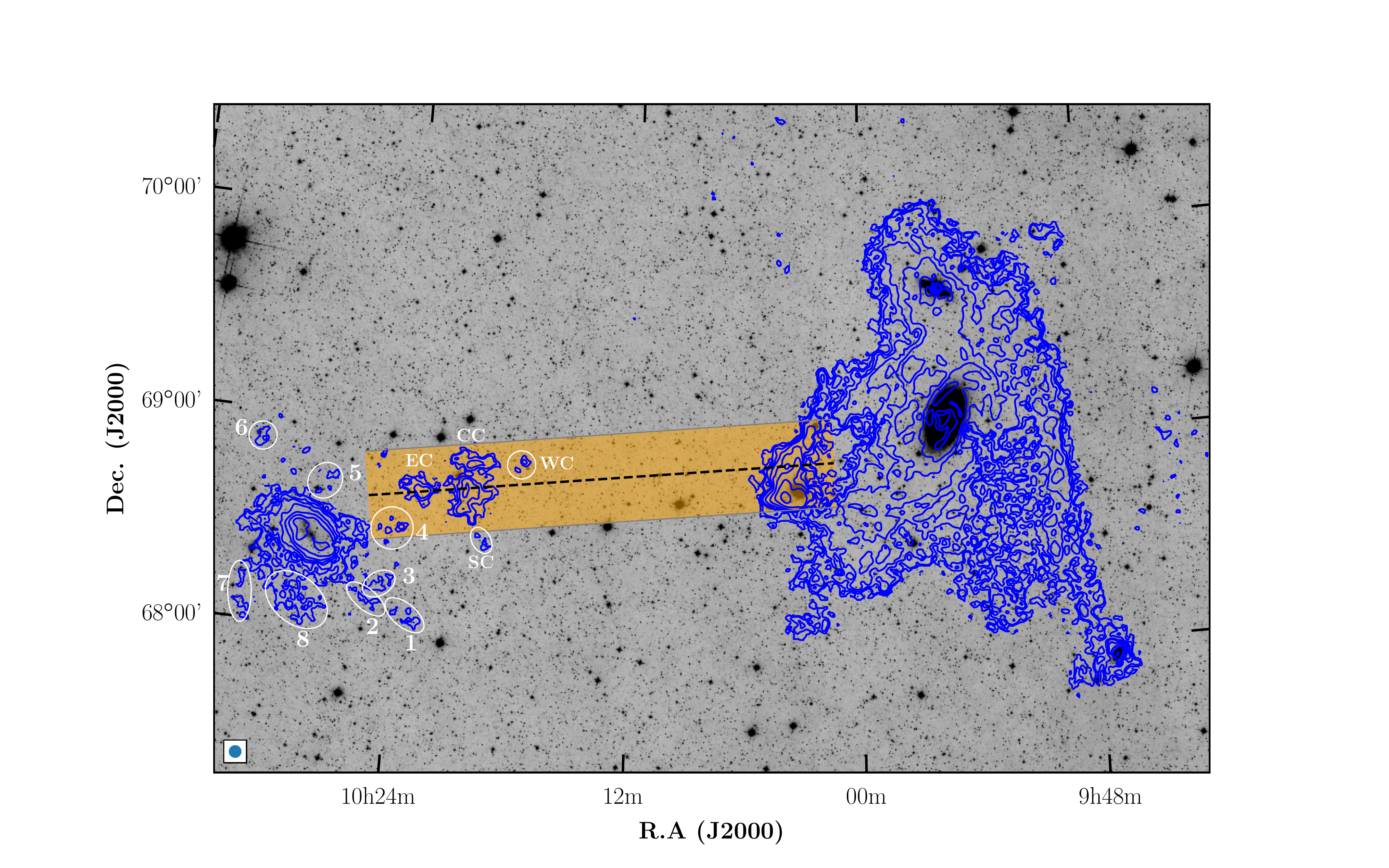}}
\vspace{-20pt}

\makebox[\textwidth][c]{
\includegraphics[width=.9\textwidth]{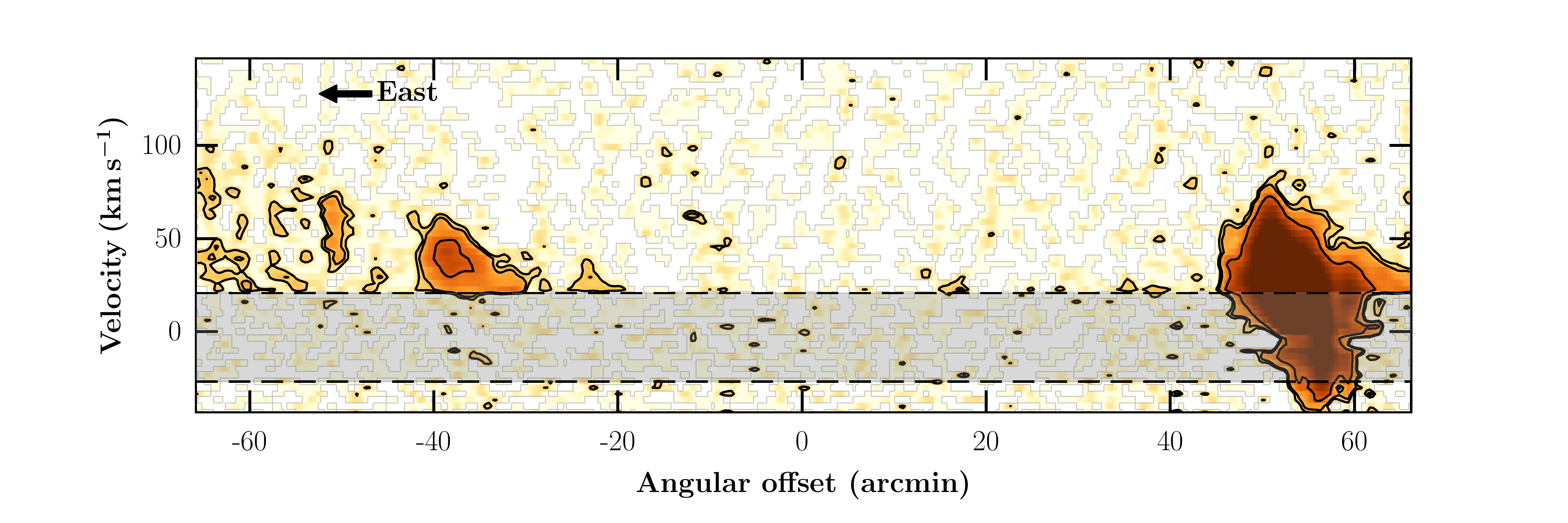}}
\vspace{-15pt}
\caption{{\it Top:} \hi\ total intensity map of the M81 region ({\it blue contours}) overlaid on a {\it WISE} W1 greyscale image. The labelled ellipses show the \hi\ clouds surrounding IC 2574, whose \hi\ masses are given in \Cref{tb:clouds}. The orange band shows the orientation of the position-velocity (PV) slice of the diagram in the bottom panel, the orange shaded area indicating the width of the slice ($22.5'$). Contours are $1, 2, 4,..., 128\times 10^{19}\,\cm$.  The blue circle at the bottom left corner of this and all subsequent figures where this appears corresponds to the size of the synthesised beam ($1.8'$) of the data. {\it Bottom:} PV diagram extracted along the dashed line of the top panel. The structures on the left of the diagram show the HIJASS J1021+68 complex of gas while the concentration on the right shows part of NGC 3077. The contours show the 2, 3, and $5\sigma$ levels.}\label{fig:m81}
\end{figure*}

In the top panel of \Cref{fig:m81} we present a total \hi\ map of the entire M81 region observed. This is an updated version of the map in Fig. 6 of Paper I, including the new observations in the region of IC 2574. Numerous clouds are detected in the neighbourhood of IC 2574 and around M81, but nothing seems to exist in between the two systems. To further verify that, we took a position-velocity (PV) slice in the space between, and connecting HIJASS J1021+68 to NGC 3077. The bottom panel of the figure shows the corresponding PV diagram, in which no apparent cloud is seen except for a few structures detected at a low column density level. These faint structures, mostly only detected above the $2\sigma$ flux level, are rather scarce in the region and do not present extended \hi\ distributions.

As shown in \Cref{tb:clouds}, the \hi\ clouds surrounding IC 2574 are low-mass objects, with masses between $10^6$ and $10^7\,\Mo$. These clouds are mostly complexes of discrete clumps, whose main members' coordinates are given in the \Cref{tb:clouds}. Cloud 8 is an extension of IC 2574's extended \hi\ envelope, with an \hi\ mass of $1.5\e{7}\,\Mo$. The table also contains an update on the \hi\ mass of the different clouds of HIJASS J1021+68, whose total mass amounts to $8.5\e{7}\,\Mo$. This is more than twice what was measured earlier in Paper I, and nearly thrice the quoted HIJASS value \citep{Boyce2001}. The properties of the \hi\ clouds surrounding the central M81 members were given in Paper I, and since the sensitivity of the data in this region has not changed, we will not discuss these here.

\begin{table}
\begin{tabular}{ccccc}
\hline
\hline
Cloud ID & RA (J2000) & Dec (J2000) & $S_{\HI}\,(\rm Jy\,km\,s^{-1})$ & $M_{\HI}\,(\rm 10^7\,\Mo)$ \\
\hline
EC & 10:23:06.2 & 68:41:00.9 & 4.59 & 1.73\\
CC & 10:20:24.9 & 68:41:50.2 & 16.8 & 6.34\\
SC & 10:19:42.3 & 68:27:22.0 & 0.23 & 0.09\\
WC & 10:17:51.2 & 68:49:27.4 & 0.36 & 0.36\\
1 & 10:23:20.4 & 68:04:37.2 & 0.61 & 0.23\\
2 & 10:25:21.9 & 68:08:17.8 & 0.41 & 0.15\\
3 & 10:24:46.4 & 68:12:59.1 & 0.70 & 0.26\\
4 & 10:24:20.4 & 68:29:00.5 & 0.59 & 0.22\\
5 & 10:28:00.5 & 68:40:46.3 & 0.36 & 0.13\\
6 & 10:31:29.1 & 68:51:40.4 & 0.66 & 0.25\\
7 & 10:31:46.8 & 68:07:00.5 & 0.98 & 0.37\\
8 & 10:28:52.0 & 68:06:14.9 & 3.98 & 1.50\\
\hline
\end{tabular}
\caption{\hi\ mass of the clouds detected in the neighbourhood of IC 2574. The IDs of the clouds correspond to their labels in \Cref{fig:m81}, and EC, CC, SC and WC designate respectively the eastern, central (including northern), southern and western components of the HIJASS J1021+68 \hi\ complex. The coordinates of the clouds (except for EC and CC) in columns 2 and 3 are those of the centres of the ellipses in \Cref{fig:m81}.}\label{tb:clouds}
\end{table}

\subsection{\hi\ emission in IC 2574}\label{sec:hi-emission}
In the following we consider a subcube, only including the region containing IC 2574 and HIJASS J1021+68. The benefit of this is that it not only allows a faster processing of the data, but it also allows one to focus on the features associated with the two objects. The global \hi\ profile of IC 2574, computed by integrating the \hi\ emission inside the \hi\ radius (see below) of the galaxy, is presented in \Cref{fig:hiprofile}. The total \hi\ line flux measured for the galaxy is $367\pm21$ Jy \kms, which corresponds to an \hi\ mass of $(1.4\pm0.1)\e{9}\,\Mo$. For comparison, the \hi\ flux of the galaxy published in \citet{Martimbeau1994} and derived using the WSRT (Westerbork Synthesis Radio Telescope) is 286 Jy \kms, i.e, ${\sim}22\%$ lower than the flux derived in the present work. The relatively high \hi\ mass of IC 2574 makes the dwarf a gas rich galaxy, with a gas fraction of $3.5\pm0.1$. The systemic velocity of the galaxy derived from the \hi\ profile is $49.5\pm0.8$ \kms, with a velocity width measured at 50\% of the peak intensity of $106.7\pm5.3$ \kms. The profile shows that the galaxy is quite asymmetric, with more \hi\ flux residing on its approaching (southwestern) side than on the receding (northeastern) side. The foreground Galactic \hi\ mainly affects the rising part of the approaching side, but the shape of the profile shows that the Milky Way subtraction procedure successfully removed most of the contamination. In \Cref{fig:hidensity} we show the \hi\ surface density profile of the galaxy, from which we derive an \hi\ diameter of $20.8'$ (or 24.4 kpc) at the $1\,\Mo\rm\,pc^{-2}$ isophote.

We present in \Cref{fig:nhimap} the total \hi\ intensity (zeroth moment) map of IC 2574 obtained from the $1.8'$ (2.1 kpc) resolution datacube. In \Cref{fig:velfield} we show the corresponding velocity field (first moment) and the velocity dispersion (second moment) maps. These moment maps were computed with the {\sc Moment} task of {\sc Miriad} using a mask previously created at a $5\sigma$ clipping level with the smooth and clip algorithm of {\sc SoFiA} \citep{Serra2015}. We have constructed the mask such that the emission from the extended \hi\ envelope of the galaxy (see \Cref{sec:extended_emission}) is excluded, and only the gas contributing to the rotation of the galaxy is accounted for. This implies that no spatial smoothing was performed on the datacube, and only a spectral smoothing to twice the channel width of the cube was done before the final mask was derived. More explicitly, this means that the mask used to compute the moment maps was obtained by combining two masks: one created from the original datacube, and the other at twice the original velocity resolution. Consistent with the maps presented in earlier works \citep[e.g.,][]{Martimbeau1994,Walter1999,Oh2008,DeBlok2008}, the velocity maps in \Cref{fig:velfield} show non-circular and random motions in the inner regions of the galaxy. These are likely the effects of the dynamics of the expanding \hi\ shells and holes residing in the galaxy \citep{Walter1999}.

\begin{figure}
\makebox[\columnwidth][c]{
\hspace{-20pt}
\includegraphics[width=1.0\columnwidth]{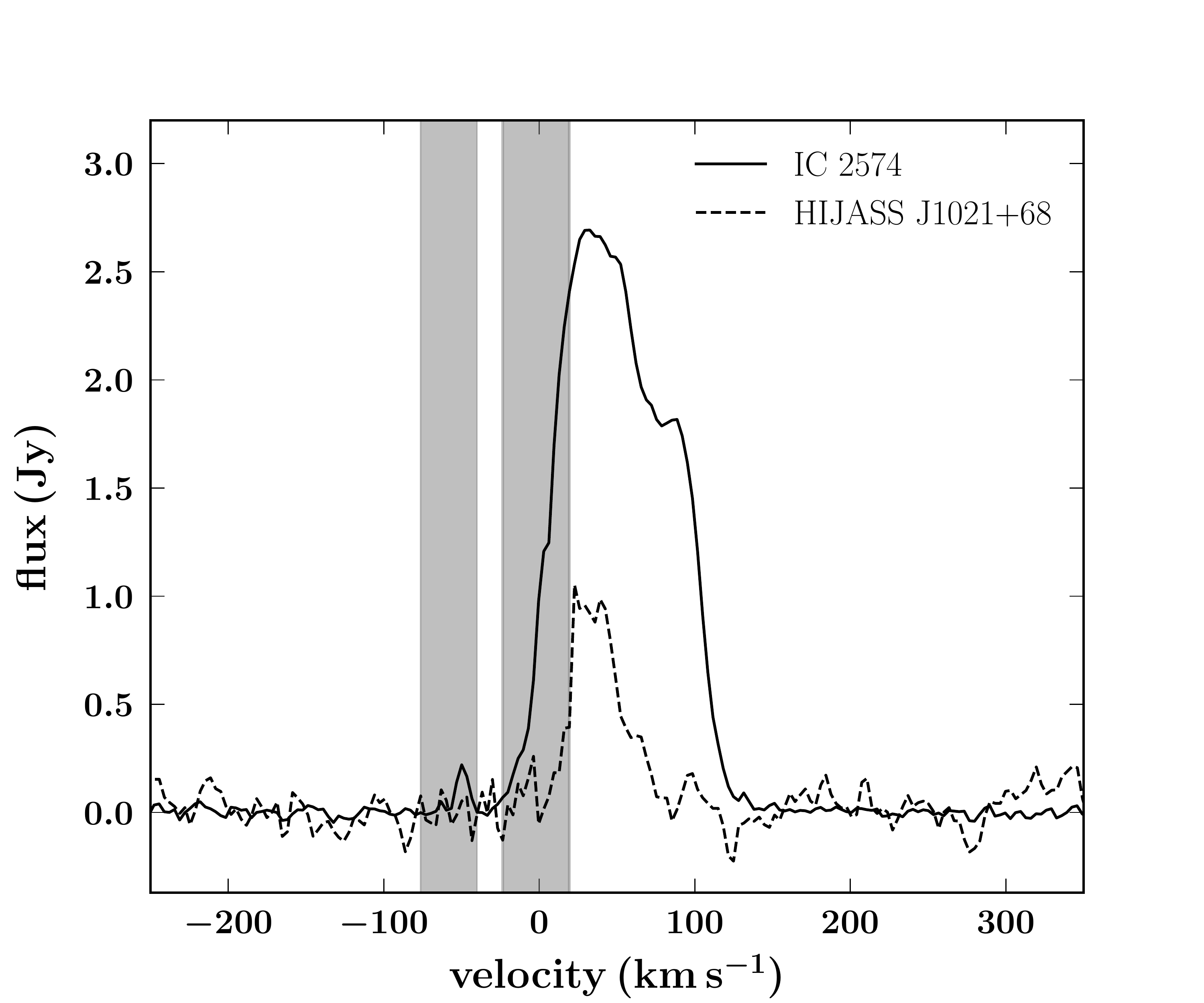}}
\vspace{-15pt}
\caption{The global \hi\ profile of IC 2574 ({\it full line}) and HIJASS J1021+21 ({\it dashed line}). The grey areas show the velocity ranges that are contaminated by foreground emission.}\label{fig:hiprofile}
\end{figure}

\begin{figure}
\makebox[\columnwidth][c]{
\includegraphics[width=\columnwidth]{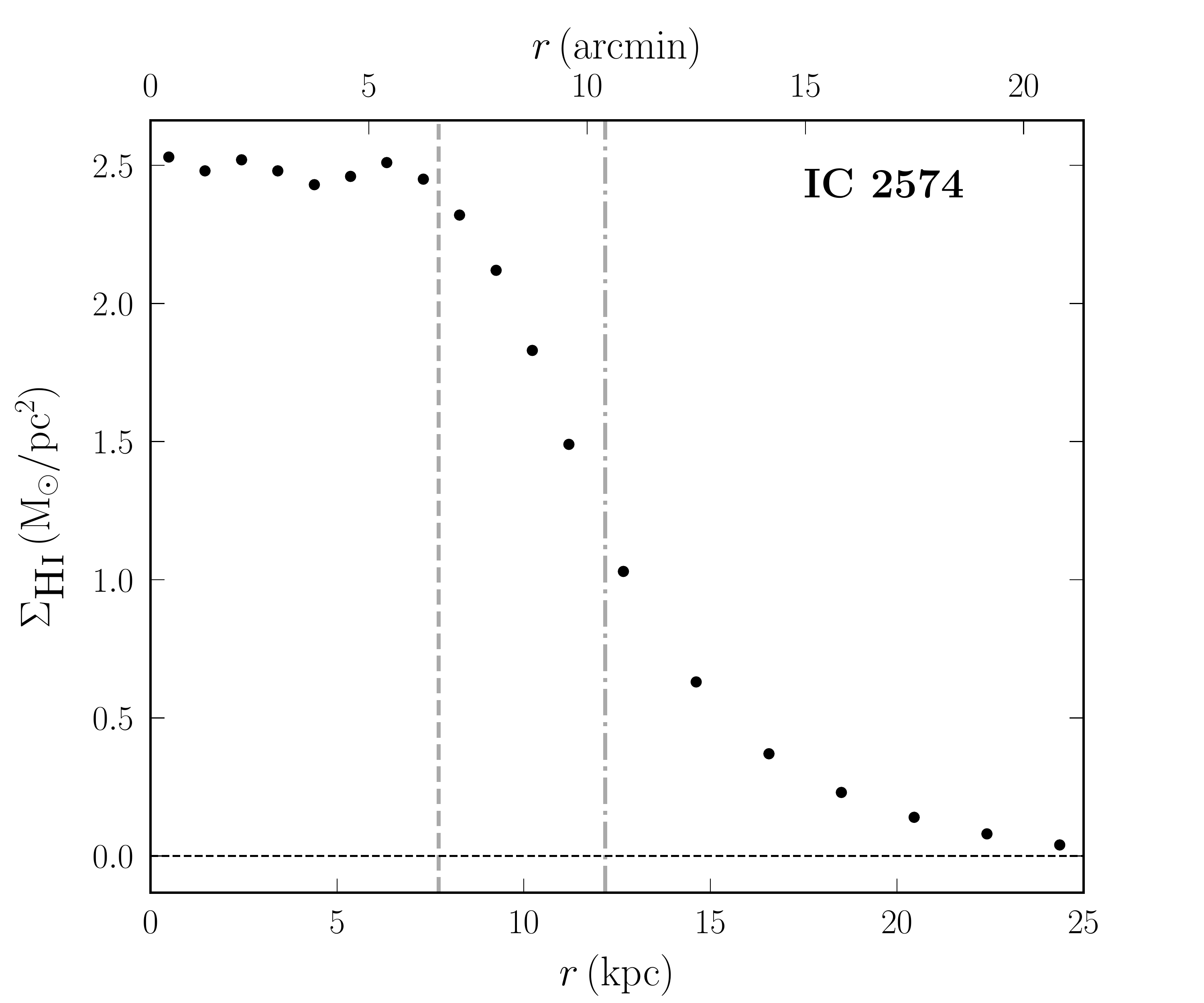}}
\vspace{-15pt}
\caption{\hi\ surface density profile of IC 2574 obtained by averaging the emission enclosed in annuli. The {\it dashed} and {\it dash-dotted} vertical lines represent respectively the optical radius measured at the 25th B magnitude and the \hi\ radius measured at $1\,\Mo\rm\,pc^{-2}$.}\label{fig:hidensity}
\end{figure}

\begin{figure}
\makebox[\columnwidth][c]{
\hspace{10pt}
\includegraphics[width=1.15\columnwidth]{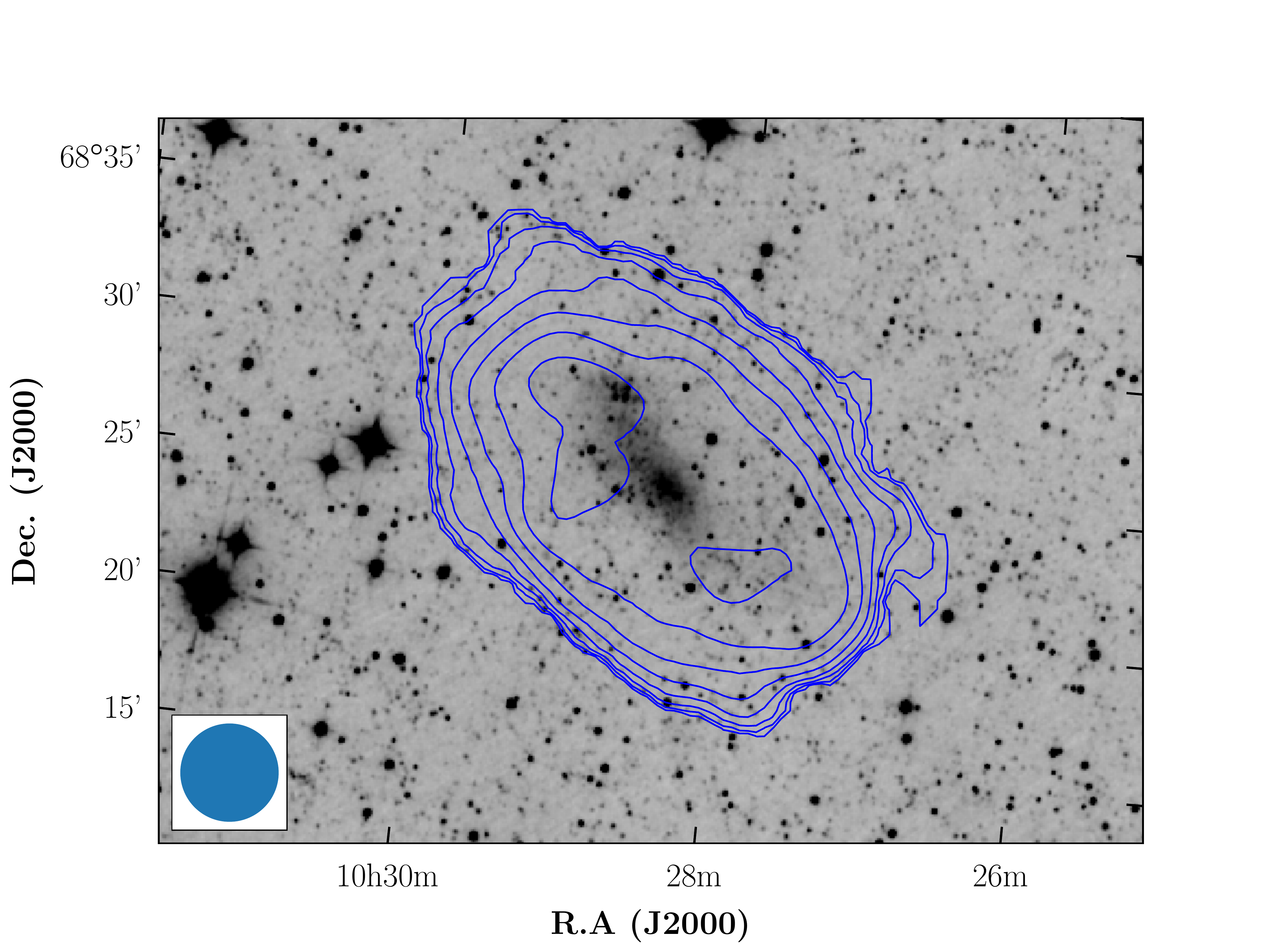}}
\vspace{-15pt}
\caption{Column density map of IC 2574 overlaid on a {\it WISE} W1 grayscale image. Contours are $1, 2, 4, \dots, 128\e{19}\,\cm$.}\label{fig:nhimap}
\end{figure}

\begin{figure*}
\makebox[\textwidth][c]{
\hspace{10pt}
\includegraphics[width=1.1\textwidth]{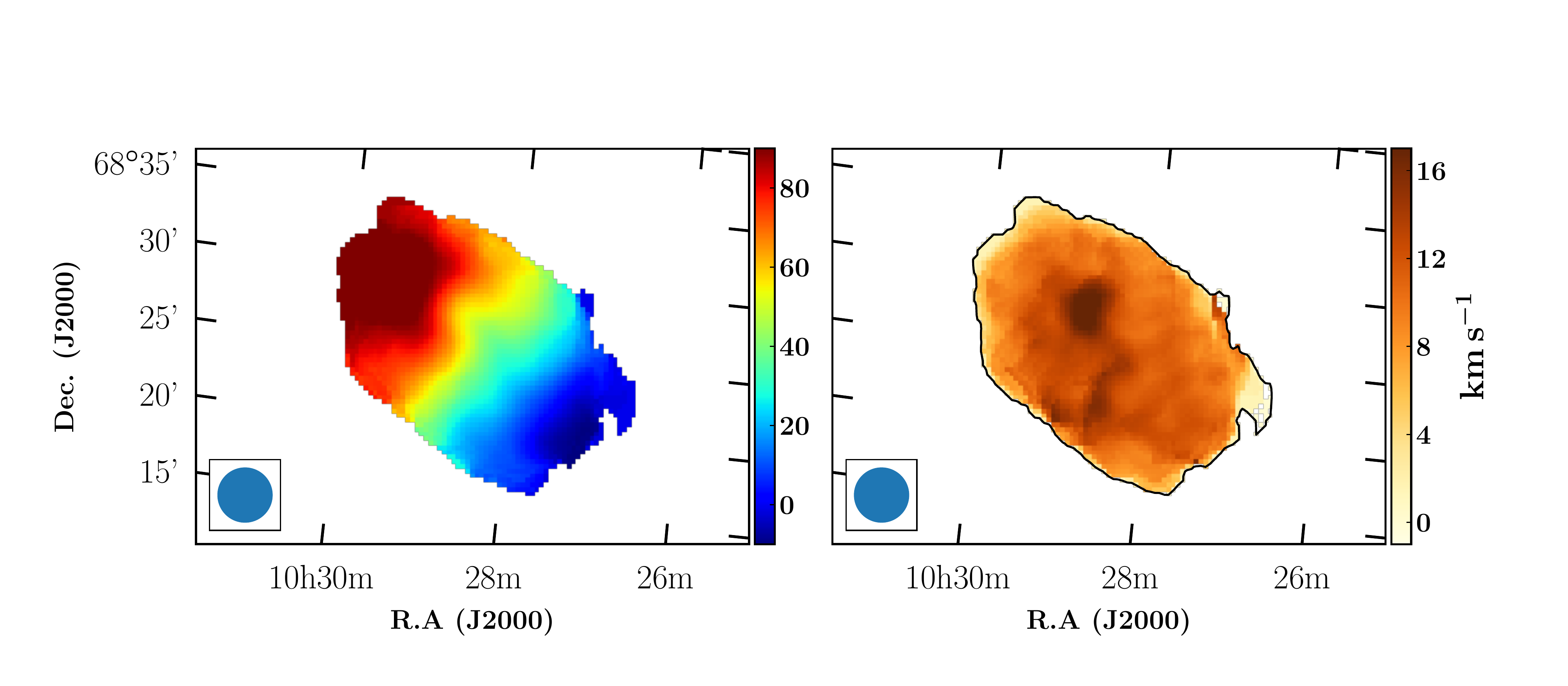}}
\vspace{-20pt}
\caption{Heliocentric velocity ({\it left}) and velocity dispersion ({\it right}) maps of IC 2574.}\label{fig:velfield}
\end{figure*}

\subsection{Extended \hi\ emission around IC 2574}\label{sec:extended_emission}
\begin{figure*}
\makebox[\textwidth][c]{
\includegraphics[width=1\textwidth]{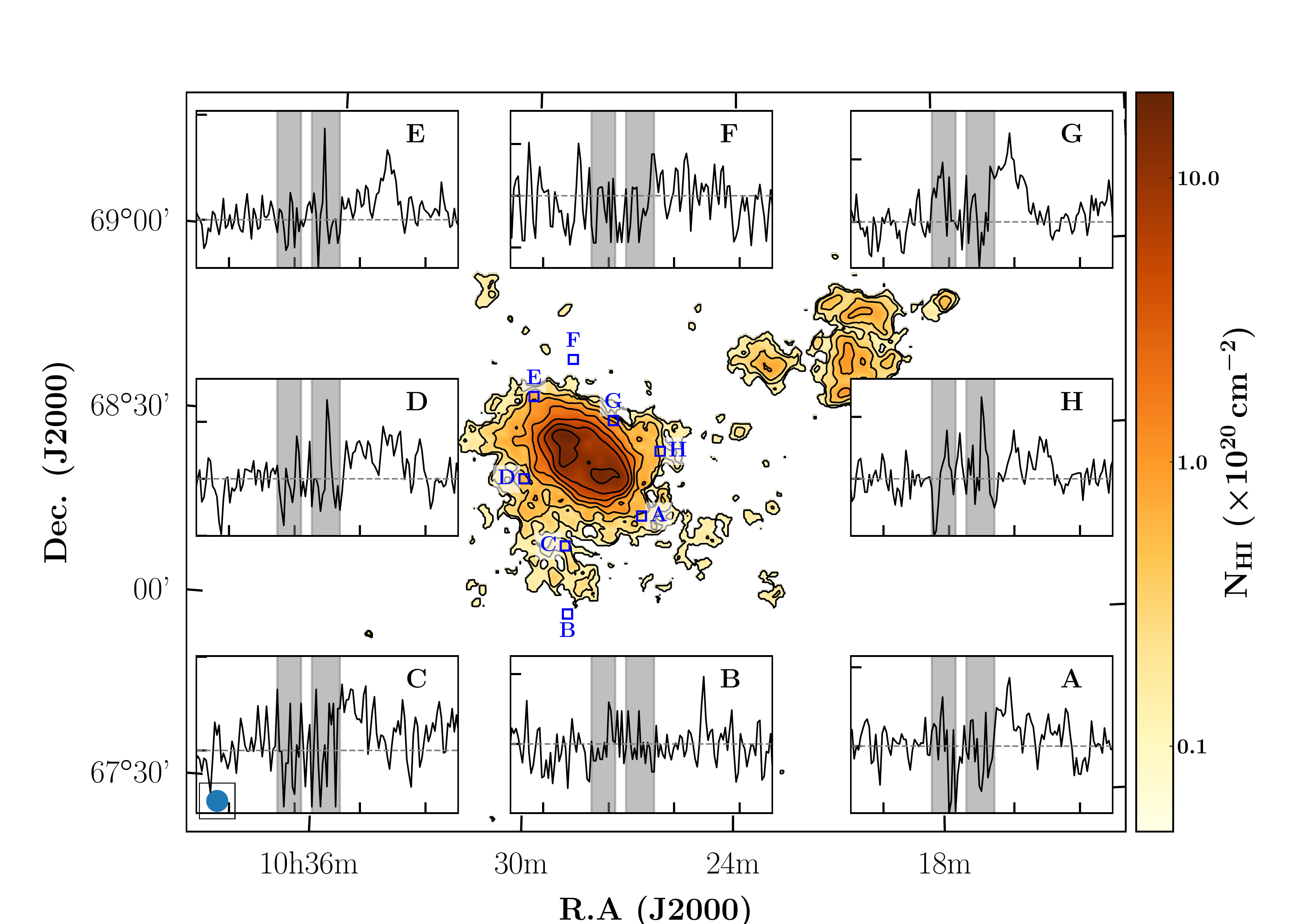}}
\vspace{-10pt}
\caption{The extended \hi\ envelope around IC 2574. The plots in the subpanels A, B, ..., H are the spectra (flux vs. velocity) extracted in and outside the \hi\ envelope at the positions delimited by boxes of corresponding names. The x-axis ticks of the plots are -150, -50, 50 and 150 \kms. The grey area shows the velocity range affected and processed for Galactic emission. The contours are $1, 2, 4, \dots, 128\e{19}\,\cm$.}\label{fig:envelope_spectra}
\end{figure*}

\begin{figure}
\makebox[\columnwidth][c]{
\hspace{-10pt}
\includegraphics[width=1.1\columnwidth]{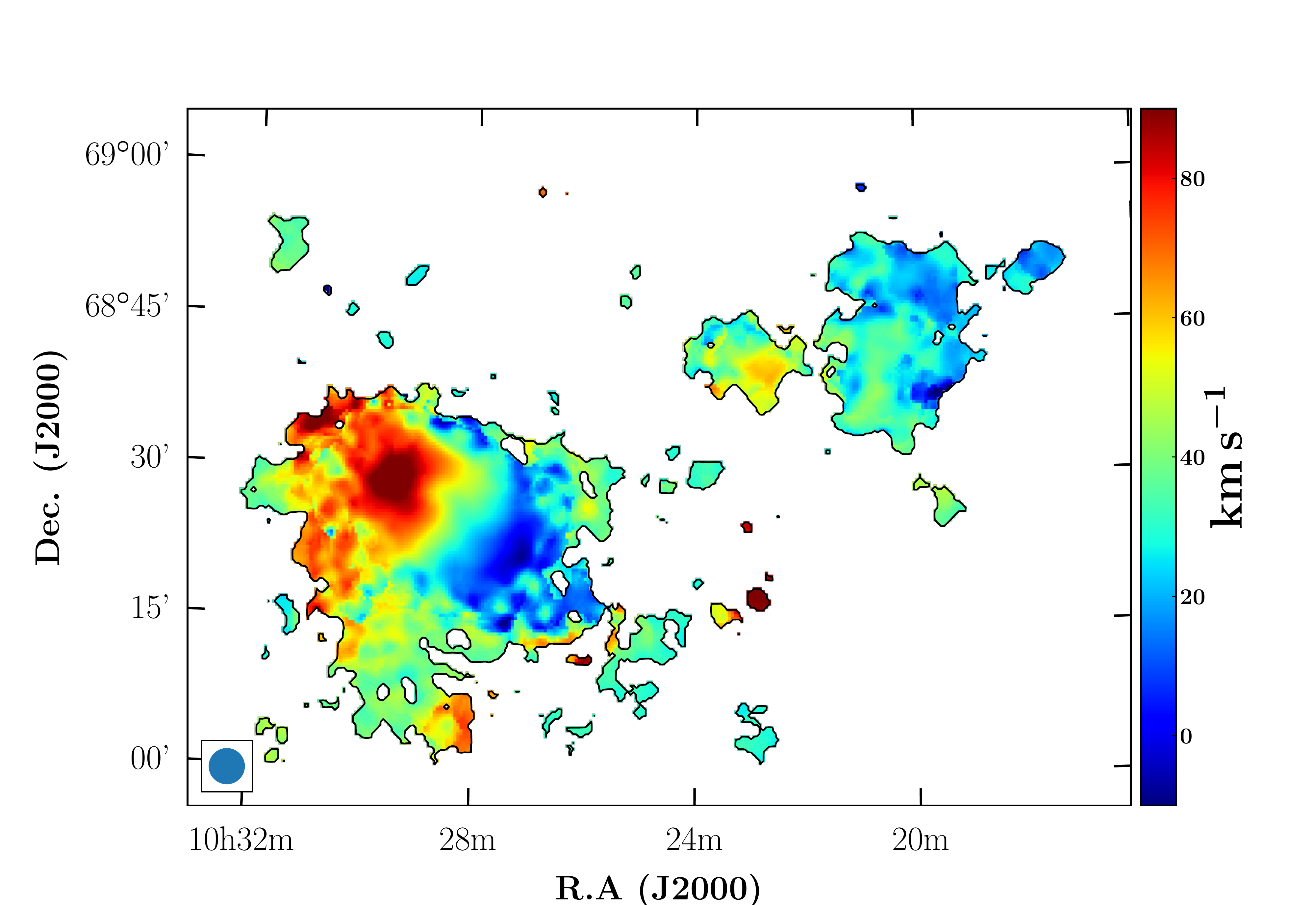}}
\vspace{-10pt}
\caption{Velocity distribution including the \hi\ envelope around IC 2574.}\label{fig:ic2574_ext_vel}
\end{figure}

To get a full view of the extended \hi\ envelope of IC 2574, we have built a mask that includes the faint emission around the galaxy. This was done by lowering the detection threshold to $4\sigma$ and expanding the number of smoothing kernels used to build the mask. The corresponding \hi\ intensity map is presented in \Cref{fig:envelope_spectra}. To ensure that the emission in the envelope is not just an artefact or foreground emission from the Milky Way, we have extracted spectra in boxes taken inside and outside the envelope (see \Cref{fig:envelope_spectra}). The spectra show that: i) most of the emission is located outside the velocity channels affected by Galactic emission, and ii) a comparison with spectra outside the envelope shows that such emission is not seen everywhere around the galaxy. 
Furthermore, the velocity distribution in the envelope (\Cref{fig:ic2574_ext_vel}) suggests that most of the low density gas, especially  the southern and western concentrations, is broadly rotating with the galaxy. The extension to the south is more or less aligned with the minor axis of the galaxy, and the gas located in that region is at systemic velocities; the gas along the western edge sits on the approaching side of the galaxy, although there appears to be a gradient in its velocity distribution. This further supports the idea that the low density gas detected around IC 2574 forms an extended envelope around the galaxy.

\subsection{\hi\ distribution in HIJASS J1021+68}\label{sec:hijass_hidist}
\begin{figure*}
\makebox[\textwidth][c]{
\includegraphics[width=1.\textwidth]{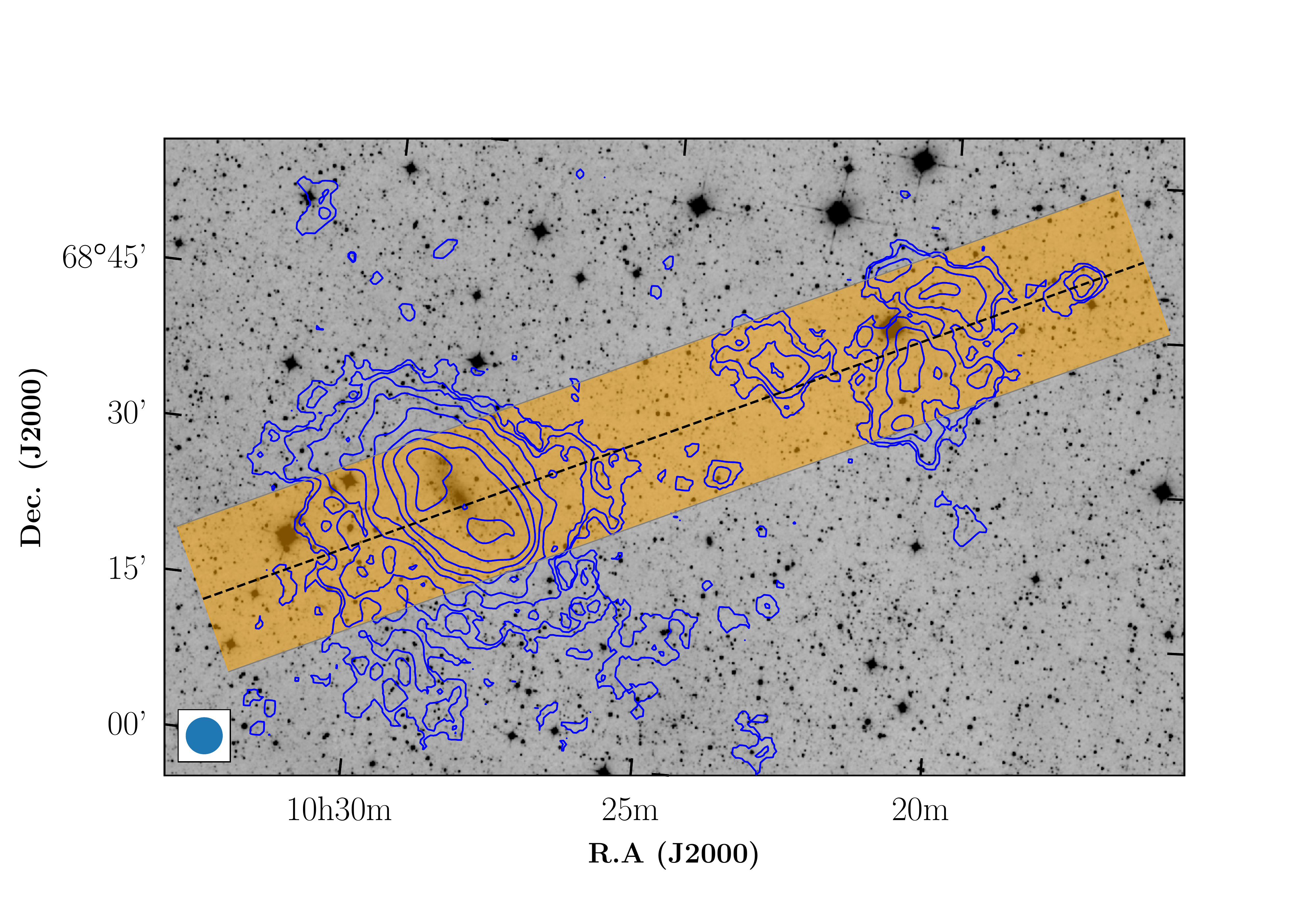}}
\vspace{-20pt}

\makebox[\columnwidth][c]{
\includegraphics[width=1.\textwidth]{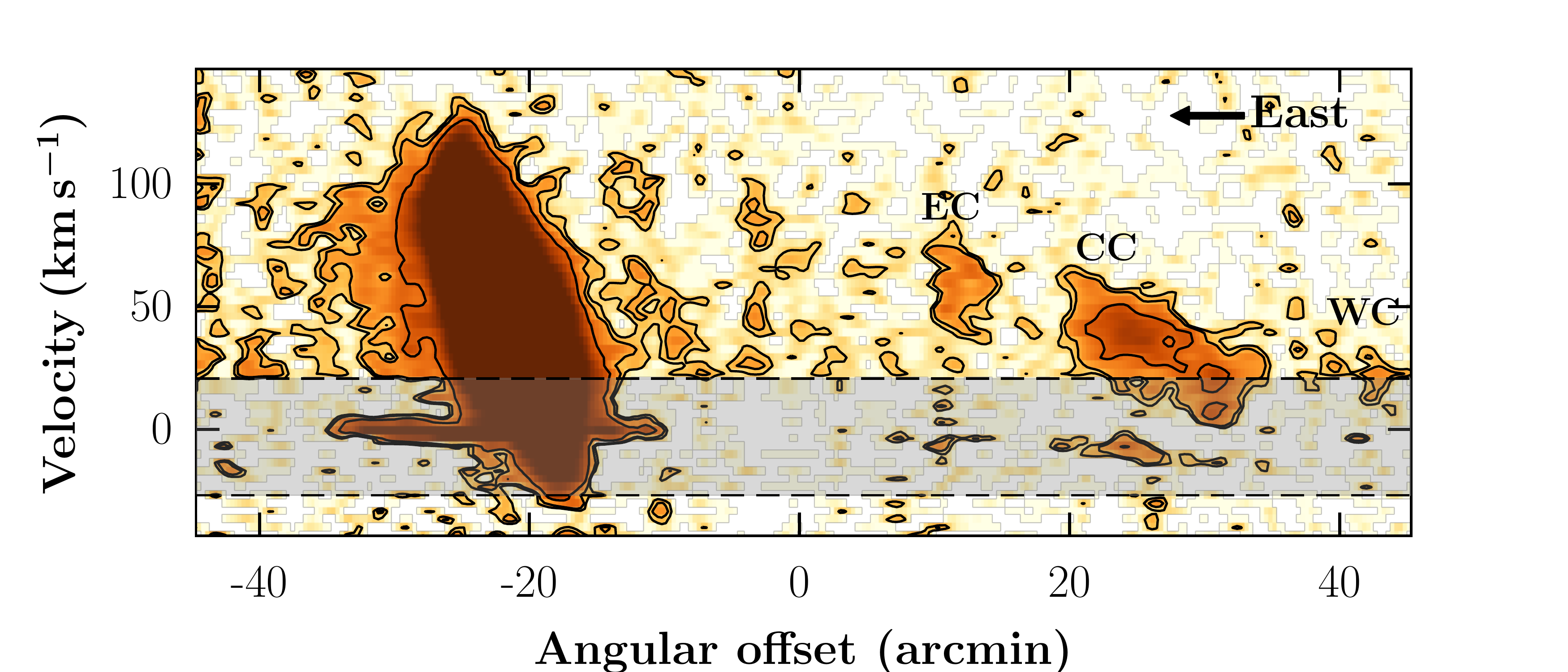}}
\vspace{-0pt}
\caption{{\it Top:} Total \hi\ intensity map of the region covering IC 2574 and HIJASS J1021+68 ({\it blue contours}) overlaid on a {\it WISE} W1 ($3.4\mu m$) grayscale image. The {\it dashed line} shows the position-velocity slice of the diagram in the bottom panel (see \Cref{sec:hijass_kinematics}) and the {\it orange shaded area} the width of the slice ($15'$). The contours are $1,2,4, ..., 128\e{19}\,\cm$. {\it Bottom:} PV diagram of the region including IC 2574 (eastern structure) and HIJASS J1021+68 (the two westernmost concentrations). There exist several small blobs in the space between IC 2574 and HIJASS J1021+68 hinting at a real connection between the two bodies. The contours represent the $2\sigma$, $3\sigma$ and $5\sigma$ flux levels, and the grey horizontal band delimited by two horizontal lines shows the velocity range that was affected by Galactic foreground \hi\ emission. The labels EC, CC and WC respectively denote the eastern, central (including the northern) and western clouds of HIJASS J1021+68. The southern cloud of the HIJASS complex is not covered by the slice used to obtain the PV diagram, and is therefore not included in the diagram.}\label{fig:ic_hijass_nhi}
\end{figure*}

Located at approximately 54 kpc west of IC 2574, HIJASS J1021+68 is a complex of gas clouds previously identified as a dark galaxy in the HIJASS catalogue \citep{Boyce2001}. The complex is not much studied in the literature, and no stellar counterpart has been reported. \Cref{fig:ic_hijass_nhi} shows an \hi\ map of the region containing IC 2574 and the gas complex. HIJASS J1021+68 appears to be a complex of two major clouds (denoted CC and EC in \Cref{fig:m81}) and two minor clouds (SC and WC), with the central cloud (CC) being the largest component. Its total \hi\ mass, as derived in \Cref{sec:m81-emission}, is $8.5\e{7}\,\Mo$. Its global \hi\ profile is shown in \Cref{fig:hiprofile}, computed by integrating the emission from the whole complex. The systemic velocity and velocity width at 50\% of the peak intensity of HIJASS J1021+68 (including all components), derived from the global profile, are respectively $35.4\pm2.9$ \kms\ and $30.8\pm5.8$ \kms. This shows that HIJASS J1021+68 has very little rotation (unless it is seen fortuitously almost face-on).

\begin{figure*}
\makebox[\textwidth][c]{
\includegraphics[width=1.1\textwidth]{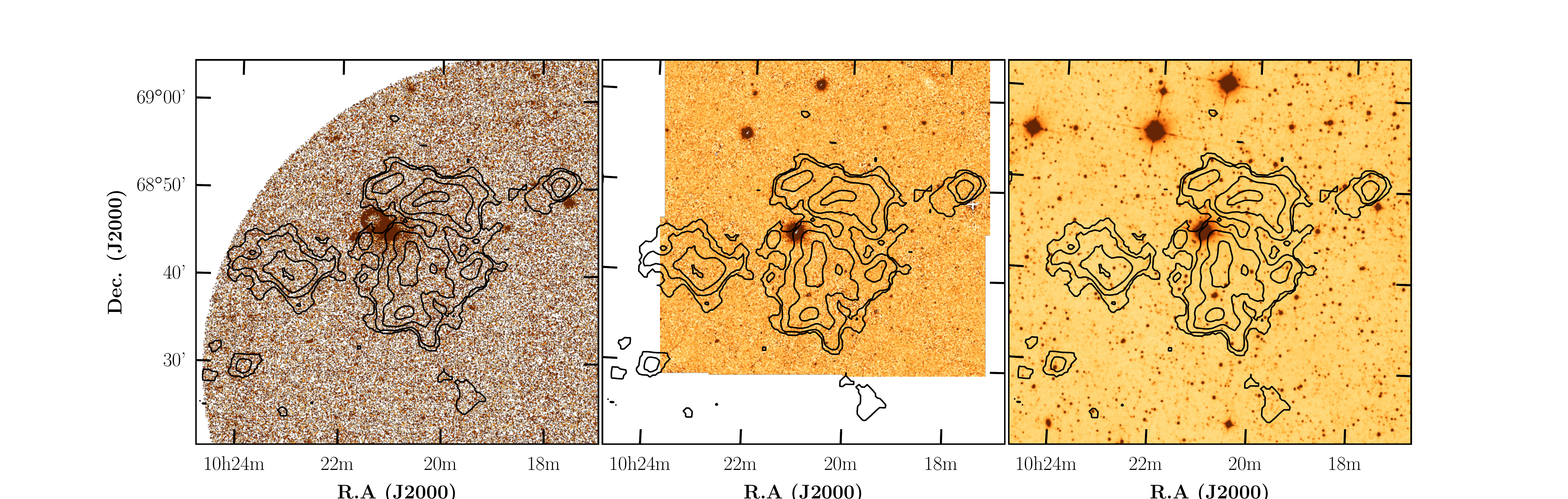}}
\vspace{-10pt}
\caption{Total \hi\ intensity map of HIJASS J1021+68 ({\it black contours}) overlaid on GALEX UV ({\it left panel}), optical R-band ({\it middle panel}) and infrared {\it WISE} ({\it right panel}) images. The contours are $1,2,4,8\e{19}\,\cm$.}\label{fig:hijass_bands}
\end{figure*}

In a search for a possible stellar counterpart for the gas complex, we collected and looked at images of the clouds in three different bands in the hope of finding a star forming region or an associated stellar population. \Cref{fig:hijass_bands} shows the \hi\ intensity map of the complex, overlaid on GALEX \citep[Galaxy Evolution Explorer;][]{Martin2005} far-UV, optical R-band, and infrared {\it WISE} \citep{Wright2010} W1 images. The R-band image is a mosaic of four different image tiles taken from the PS1 \citep[{\it aka} PanSTARRS1: Panoramic Survey Telescope and Rapid Response System;][]{Chambers2016,Flewelling2016} catalogue\footnote{Accessible via MAST: \url{http://catalogs.mast.stsci.edu/}} of images. At the spatial position of HIJASS J1021+68, none of the images present a structure that may be a possible counterpart to the complex. Considering the {\it WISE} image whose W1 band has a surface brightness noise of 23.2 \marc\, ($0.16\,\rm\mu Jy\,arcsec^{-2}$) per pixel \citep{Jarrett2012}, and considering a minimum of 1000 pixels (the angular pixel size being $2.75''$), we find that the upper limit of the noise in the image shown in the right panel of \Cref{fig:hijass_bands} is about $0.015\,\rm\mu Jy\,arcsec^{-2}$ (or 25.8 \marc). Converting this surface brightness level to luminosities yields about $0.43\,\rm\,L_\odot\,pc^{-2}$, which corresponds to an upper limit on stellar mass of $1.4\e{5}\,\Mo$ assuming a conservative mass-to-light ratio of 1 \ml.
The non-detection of stellar counterparts down to this level most likely suggests that no star formation has taken place ({\it yet}) in the gas clouds. We also note that the dwarf galaxy survey centred on M81 by \citet{Chiboucas2009} -- and follow-up in \citet{Chiboucas2013} -- with the CFHT detected no stellar counterparts at the position of HIJASS J1021+68.

\section{\hi\, kinematics}\label{sec:kinematics}

\subsection{Rotation curves of IC 2574}\label{sec:rotcur}
\subsubsection{Tilted-ring modelling}\label{sec:tiltedring}
Using the 3D tilted-ring model fitting tool \barolo\ \citep{DiTeodoro2015}, we derived the rotation curve of the galaxy. Unlike the traditionally-used \rotcur\ \citep{Begeman1989} task in \gipsy\ \citep{vanderHulst1992} which derives the rotation curve from the two-dimensional velocity field, \barolo\ takes as input the three-dimensional \hi\ datacube and fits a 3D tilted-ring model to it. In fact, because of the relatively low spatial resolution of the data (only $\sim6$ beams across IC 2574), rotation curves derived from low-resolution velocity fields are generally sensitive to beam smearing effects, which tend to underestimate the velocities in the inner regions of galaxies \citep[e.g.][]{Swaters2009,Lelli2010}. However, the 3D fitting procedure was shown to be less sensitive to these effects, and describes more accurately the rising parts of rotation curves \citep[e.g.,][]{DiTeodoro2015,Sorgho2019}.

We supplied to the fitting program the MW-subtracted \hi\ datacube of IC 2574 and the $5\sigma$ mask derived in \Cref{sec:hi-emission} with SoFiA, and provided the optical values of the inclination and position angle as initial guesses. As for the systemic velocity and the dynamical centre, we estimated the initial values from the velocity field in \Cref{fig:velfield}. We have run the program for the entire galaxy (both sides), allowing it to run twice: a first stage during which it allows all the kinematical parameters to vary, then performs a parameter regularisation where the parameters are interpolated, and a second stage during which these parameters (except the rotation velocity) are kept fixed to the fitted values. The rings in the tilted-ring model were chosen to be $50''$ wide (i.e, about half the beam size) in the inner regions of the galaxy, and because of the non-regular shape of the object (see \Cref{fig:velfield}), a ring width of $100''$ was chosen in the outer regions. Also, \barolo\ lets the user choose the fitting function (polynomial or Bezier) for the inclination and the position angle, and a series of tests has shown that the best results are obtained with a third order polynomial function.

The dynamical centre and systemic velocity obtained from the fit are respectively ($\alpha,\delta$) = ($10\dhr28\dm36\ds1,68^\circ25'39.1''$) and $45.4\pm0.7$ \kms. To ensure the consistency of these values, we have used \rotcur\ to re-estimate these values from the velocity field in \Cref{fig:velfield}, and obtained the sensibly similar value -- with an angular separation equivalent to one third of the synthesised beam -- of ($10\dhr28\dm43\ds3,68^\circ25'34.3''$) for the dynamical centre, but a higher value of $56.3\pm2.1$ \kms\ for the systemic velocity. This difference in systemic velocity values can however be attributed to the limited -- spatial and mostly spectral -- resolutions of the data. Indeed, the low resolution velocity field of \Cref{fig:velfield} puts little constraint on the 2-dimensional tilted-ring model, as compared to the 3-dimensional fit; and this, combined with the presence of asymmetric features such as non-circular and random motions are likely to affect the spectral position of the centre.

\begin{figure*}
\makebox[\textwidth][c]{
\hspace{30pt}
\includegraphics[width=1.1\columnwidth]{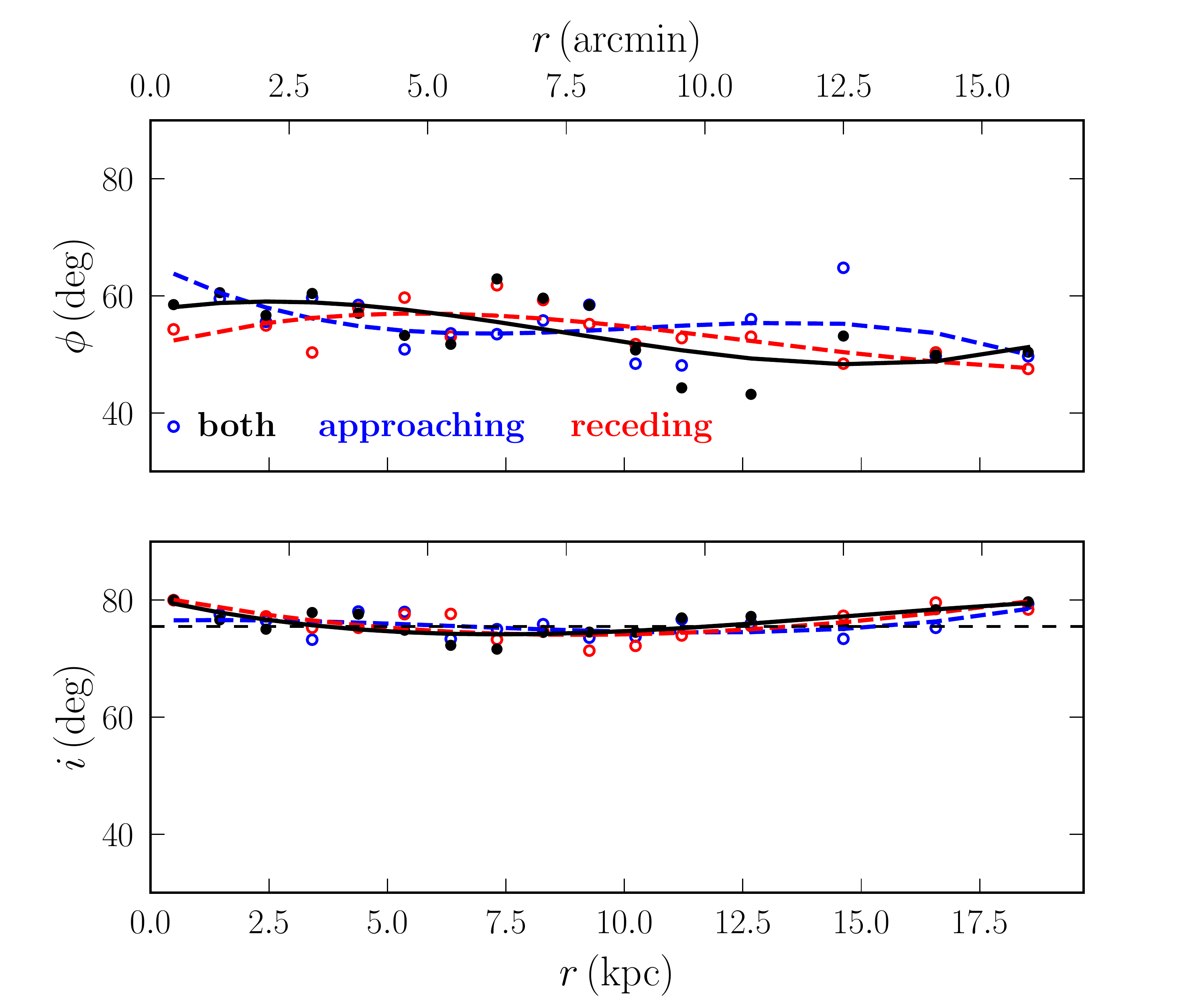}
\includegraphics[width=1.1\columnwidth]{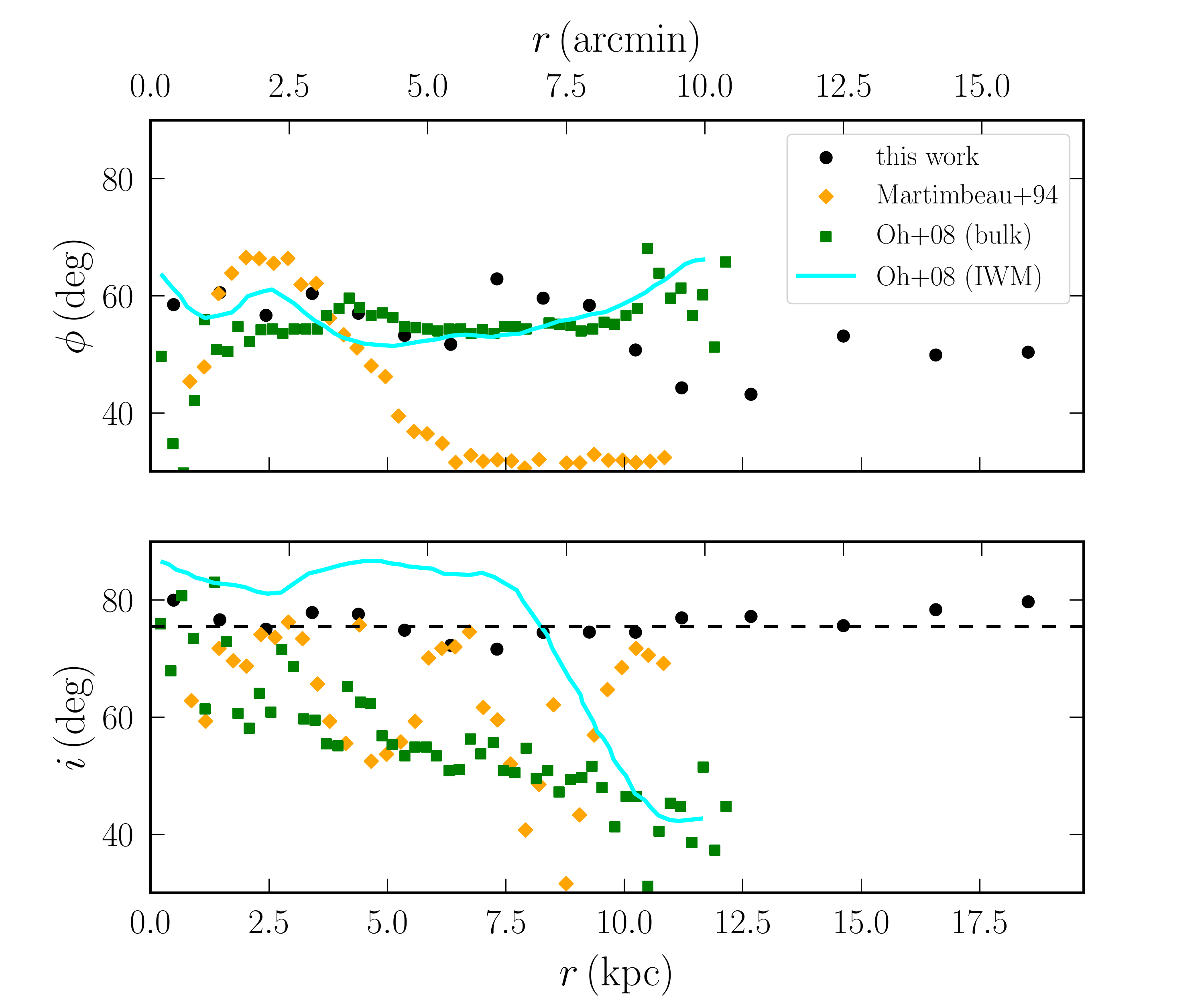}}
\vspace{-10pt}
\caption{{\it Left:} Tilted-ring model fits to position angle, $\phi$, and inclination, $i$, of IC 2574 obtained from the {\sc 3d Barolo} fitting program. The horizontal dashed line in the lower panel shows the $\rm 3.6\,\mu m$ photometric inclination of the galaxy \citep{Salo2015}. {\it Right:} comparison of the variations of $\phi$ and $i$ to the literature.}\label{fig:tiltedring}
\end{figure*}

Fixing the centre and systemic velocity of the galaxy to the above values derived with \barolo, we have run the program two more times to estimate the variation of both the position angle and inclination for the approaching and receding halves, respectively. These are shown in the left panel of \Cref{fig:tiltedring}. Except for a few minor differences, we note a good agreement between the sides and the resultant model for the variations of both the inclination and the position angle. Indeed, we see that for all the 3 models, there is a slight decrease in the variation of the position angle with radius, while the inclination tends to be high (${\sim}80\dg$) in the centre and in the outskirts of the galaxy, with the lowest values (${\sim}70\dg$) found in the region $\sim4-10$ kpc. This similarity in the variations of the models' geometric parameters suggests a kinematical symmetry in the galaxy's disc. In the right panel of \Cref{fig:tiltedring}, we compare the tilted-ring model fits to the galaxy's position angle and inclination to those found in the literature. While the variations of the position angle obtained in the present work seem to agree -- in the inner 10 kpc -- with those of the THINGS bulk and intensity-weighted mean (IWM) solutions \citep{Oh2008}, we note a clear difference with the values of the position angle derived in \citet{Martimbeau1994}. The latter work's values of $\phi$ are on average lower than those obtained in \citet{Oh2008} and this work, particularly beyond a 3 kpc radius.

As for the inclination, we note that the solutions obtained from \barolo\ sit on the upper limit of the inclinations of \citet{Martimbeau1994}, which are scattered between ${\sim}40\dg$ and ${\sim}75\dg$ throughout the probed radii. Interestingly, the inclinations of the bulk model of \citet{Oh2008} constantly decrease from the centre -- where the values are about $80\dg$ --  towards the outer regions (to about $40\dg$), while the present work's values tend to remain relatively constant. On the other hand, the inclinations of IWM's model are higher than our values in the inner 8.5 kpc of the galaxy, and abruptly fall outwards.

\begin{figure}
\makebox[\columnwidth][c]{
\hspace{10pt}
\includegraphics[width=1.15\columnwidth]{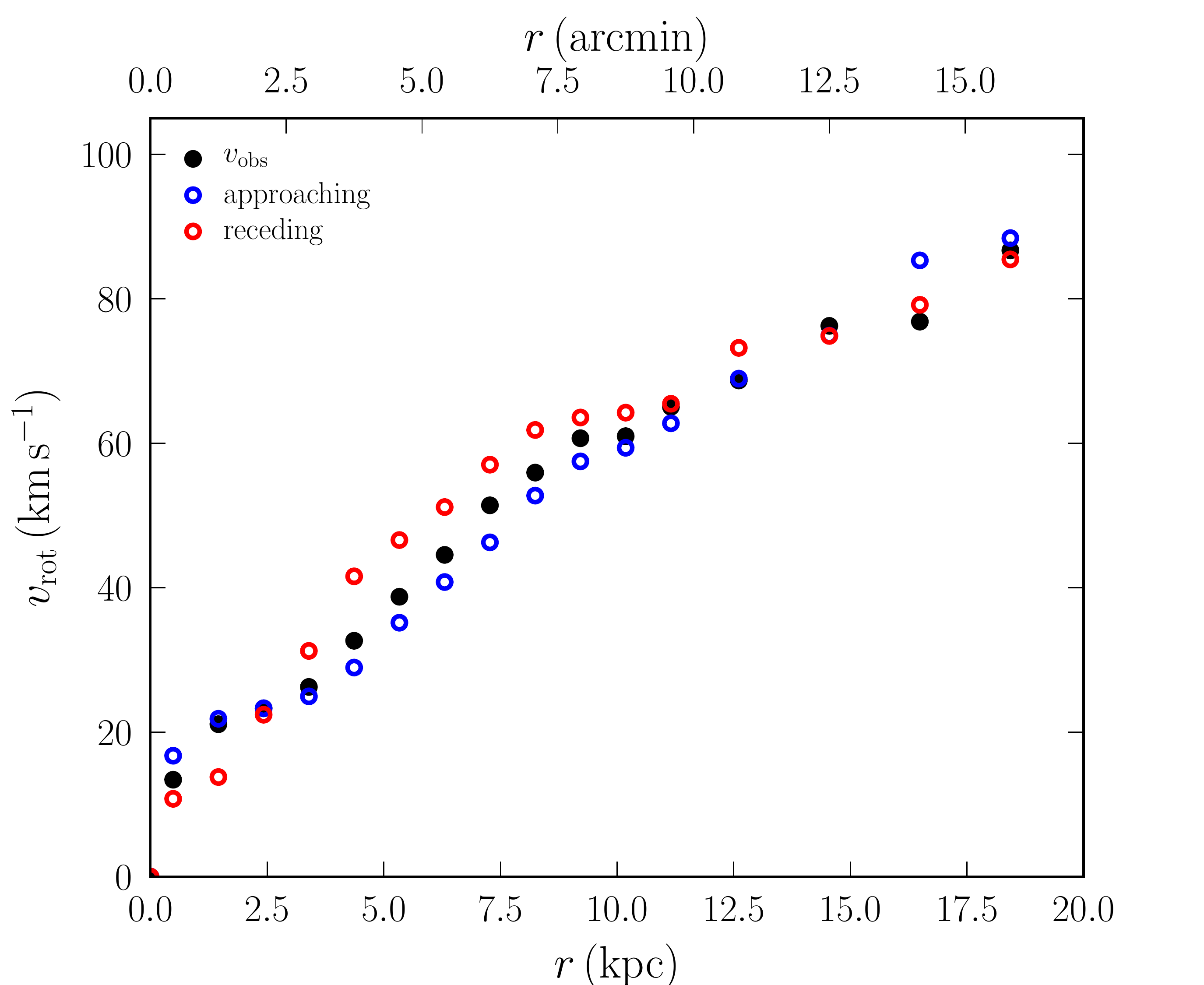}}
\vspace{-10pt}
\caption{Rotation curve of IC 2574 derived using the parameters of \Cref{fig:tiltedring}.}\label{fig:rotcur}
\end{figure}

We ran the program once more on the approaching and then on the receding half of the galaxy, with all the geometrical parameters fixed to those obtained for the overall model. The three rotation curves are presented in \Cref{fig:rotcur} where, as expected, the rotation of the galaxy seems to follow a solid-body type of rotation.

\subsubsection{Asymmetric drift correction}\label{sec:adrift}
\begin{figure}
\makebox[\columnwidth][c]{
\hspace{20pt}
\includegraphics[width=1.15\columnwidth]{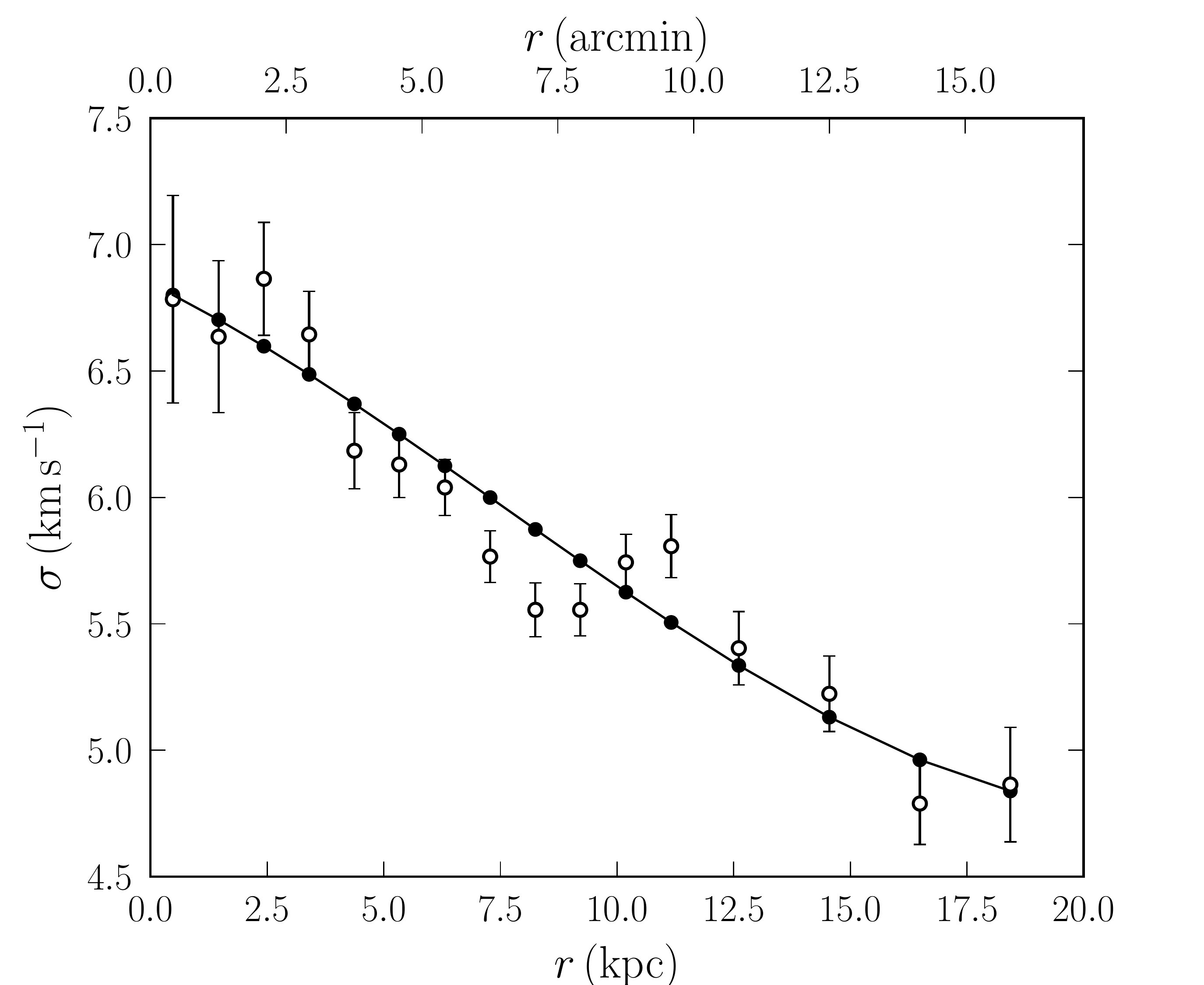}}
\vspace{-15pt}
\caption{Velocity dispersion profile of IC 2574. The open symbols represent the observed velocity dispersions while the filled symbols and the line show the {\it smoothed} velocities used to determined the asymmetric drift corrections.}\label{fig:vdisp}
\end{figure}

\begin{figure}
\makebox[\columnwidth][c]{
\hspace{20pt}
\includegraphics[width=1.15\columnwidth]{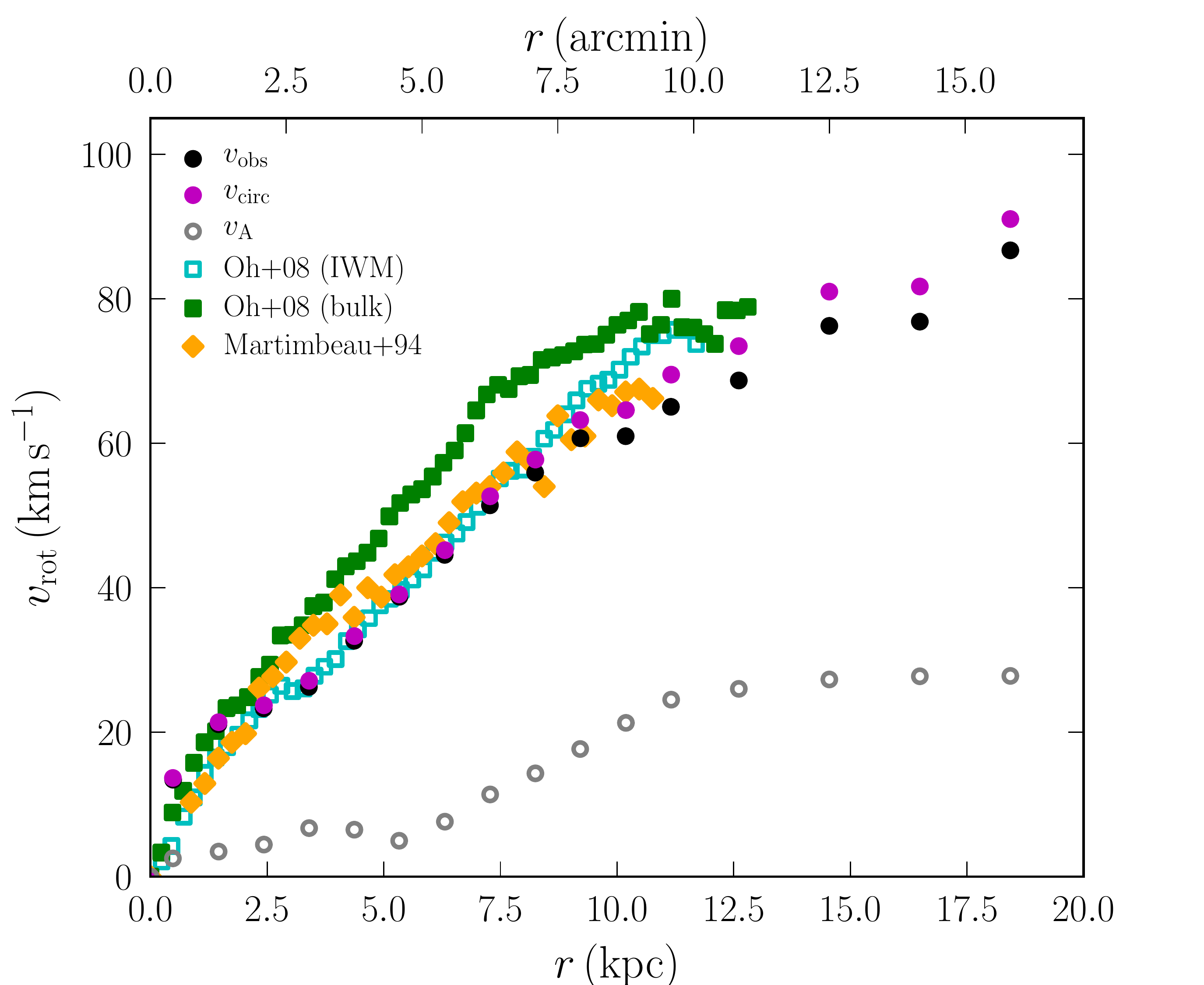}}
\vspace{-15pt}
\caption{Rotation curves of IC 2574 compared with the literature. $v_A$ represents the asymmetric drift velocity, given by the second term of the right-hand side of \Cref{eq:adrift}.}\label{fig:adrift}
\end{figure}

Because of the low mass and rotation velocity of IC 2574, one expects the rotation curve of the galaxy to be underestimated because of the effects of asymmetric drift \citep[e.g.,][]{Valenzuela2007,Dalcanton2010,Pineda2017}.
To obtain the true circular velocity (i.e, excluding the non-circular motions) of the galaxy, we use Eq. (4.227) of \citet{Binney2008} with the assumption that the product $v_r v_z$ is independent of $z$, and that the thickness of the disc is constant; we therefore simplify the above equation to \citep[also see][]{Cote2000,Carignan2013,Chemin2016}
\begin{equation}\label{eq:adrift}
v^2_{\rm circ}(r) = v^2_{\rm obs}(r) - r\,\sigma^2(r)\,\left[{d\over dr}\ln{\Sigma(r)} + {d\over dr}\ln{\sigma}^2(r)\right]
\end{equation}
where $v_{\rm obs}(r)$, $\sigma(r)$ and $\Sigma(r)$ are respectively the observed rotational velocity, the observed line-of-sight velocity dispersion and the total gas surface density at a radius $r$. The line-of-sight velocity dispersion was derived by fitting ellipses to the second moment map of the galaxy, and its variation is given in \Cref{fig:vdisp}. The velocities were corrected for instrumental broadening, and the profile was smoothed with a third order polynomial to make it less sensitive to the small-scale, local fluctuations. Following \citet{Dalcanton2010}, the total gas surface density was assumed to be 1.4 times higher than the \hi\ surface density of \Cref{fig:hidensity}, where the factor 1.4 allows to approximately correct for the helium and heavier elements. In \Cref{fig:adrift} we show a comparison of the observed rotational and circular velocities, and those derived in previous studies in the literature. We also show on the plot the asymmetric drift velocity, given by the second term of the right-hand side of \Cref{eq:adrift}.

The rotation curves derived in the present work extend farther than those in the literature, and overall agree well with the THINGS IWM \citep{Oh2008} and \citet{Martimbeau1994}'s rotation curves in the central regions of the galaxy. However, the velocities of the derived rotation curves in the inner regions of the galaxy ($<2$ kpc) are overestimated by 11\% to 27\% when compared to these higher resolution curves. In the region $\sim10-13$ kpc, the derived rotation curves are slightly below the THINGS curve, but agree with the latter within the error bars (not shown for clarity). The figure also shows the THINGS bulk rotation curve from \citet{Oh2008}, derived from the galaxy's \hi\ velocity field after correcting the data for random and small-scale non-circular motions. The bulk rotation curve is above the curves derived in this work in the region $>2.3$ kpc, but its last points agree with the circular velocities of the present work.

\begin{figure}
\makebox[\columnwidth][c]{
\includegraphics[width=1.1\columnwidth]{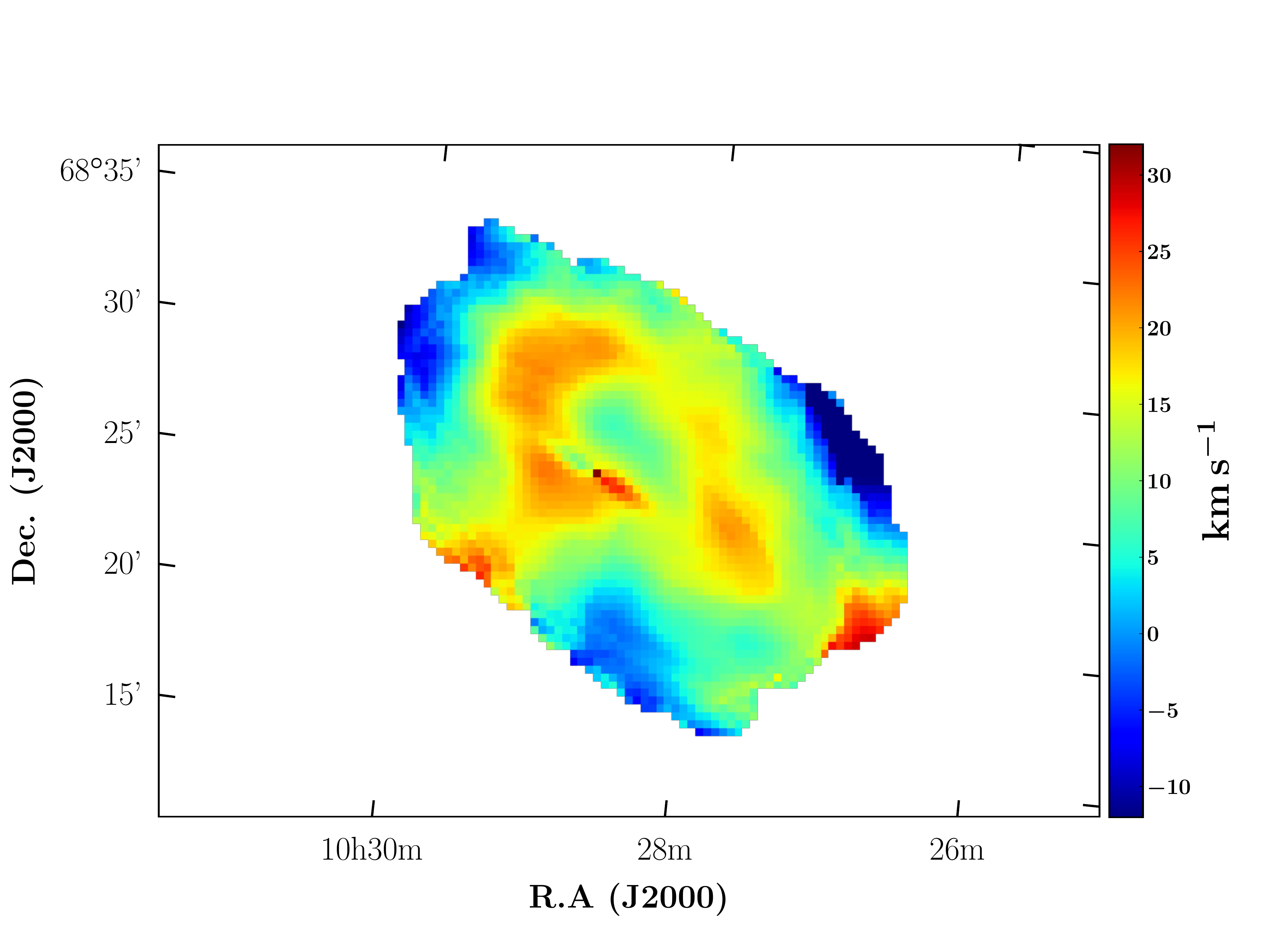}}
\vspace{-20pt}
\caption{Residual velocity field of IC 2574. The regions of high residuals around the centre show where the non-circular motions influence the rotation of the galaxy.}\label{fig:residual}
\end{figure}

To get a comparison between the tilted-ring model and the observed kinematics of the galaxy, we have constructed a residual map of the velocity field (\Cref{fig:residual}) by subtracting a model velocity field of from the observed one. The model velocity field was computed using the \barolo\ geometrical parameters and the circular rotation curve obtained after accounting for the asymmetric drift. The residual map shows the regions of the galaxy where the non-circular motions mostly affect the rotation of the galaxy, with a difference velocity between the model and the data of the order of about $10-25$ \kms. It also shows that the residual velocities are highest in the outermost regions of the galaxy, especially around the southwestern side.

\subsection{Mass model}\label{sec:massmodel}
To construct the mass model of IC 2574, we take advantage of both the high resolution VLA data and the extended DRAO curve: we construct a {\it hybrid} rotation curve by combining the bulk velocity rotation curve of \citet{Oh2008} to that of the circular velocity derived in this work and plotted in \Cref{fig:adrift}. In the inner regions ($r<13$ kpc) we consider the high resolution curve only, while in the outer regions we consider the lower resolution DRAO curve.
Also, because the THINGS curve is oversampled (two points per beam), we resampled it by dropping every second point in the curve, following the mass models of THINGS rotation curves in \citet{Chemin2011}.

\subsubsection{Baryonic component}
\begin{figure}
\makebox[\columnwidth][c]{
\includegraphics[width=1.\columnwidth]{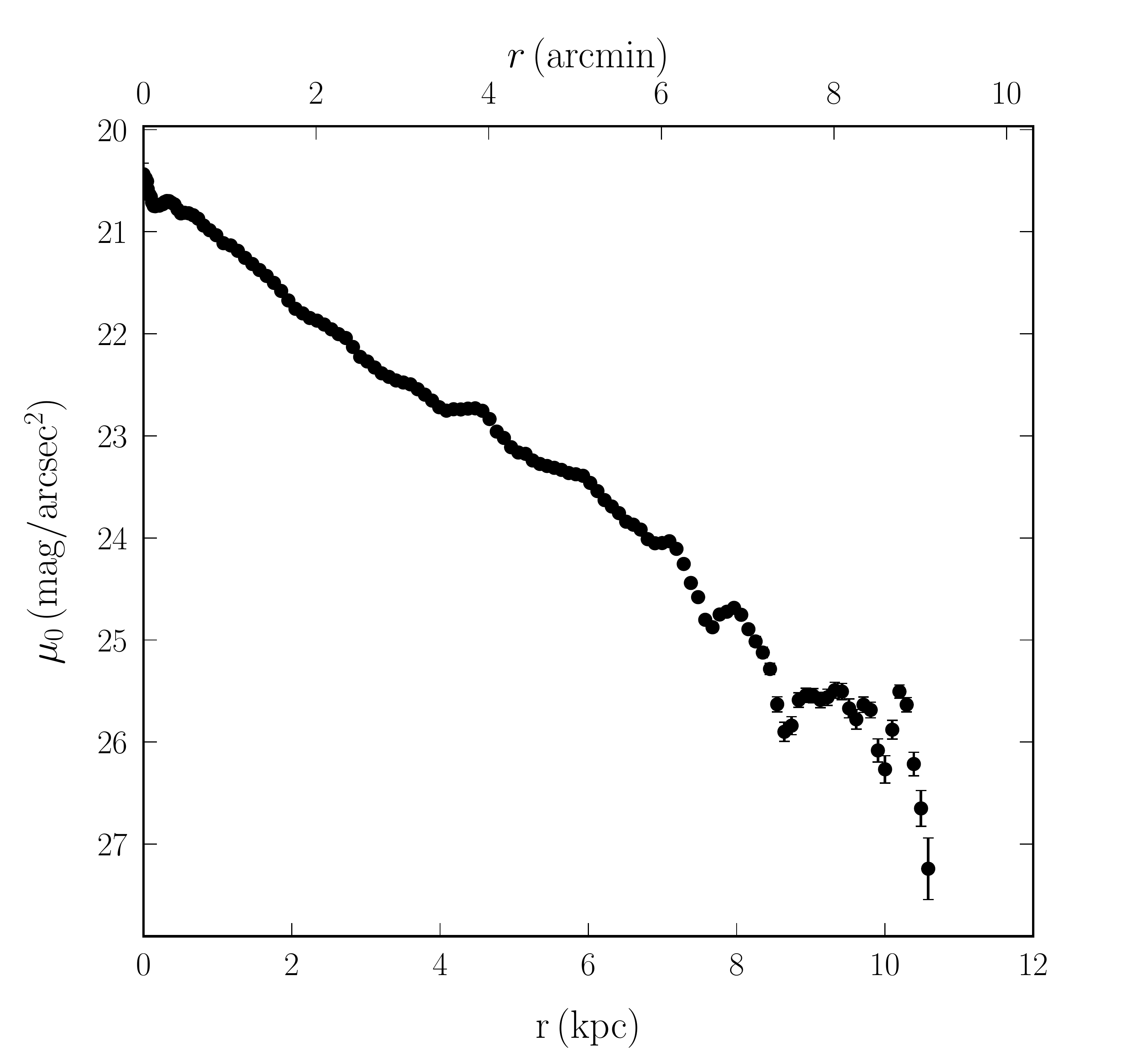}}
\vspace{-20pt}
\caption{W1 ($\rm 3.4\,\mu m$) surface brightness profile of IC 2574.}\label{fig:w1profile}
\end{figure}

The {\it WISE} \citep{Wright2010} W1 ($3.4\mu m$) light profile of IC 2574 (from Jarrett et al., {\it in prep.}) is shown in \Cref{fig:w1profile}. The profile suggests the absence of a prominent bulge as expected, and an attempt to perform a decomposition into different components -- an exponential disc and a spherical bulge -- further confirmed that the bulge contribution to the total luminosity of the galaxy is negligible, with a bulge-to-total luminosity ratio of only 6\%. We therefore assume that the only stellar component of the galaxy is the stellar disc.

For the gaseous disc component, we derived the contribution from the galaxy's \hi\ surface density presented in \Cref{fig:hidensity}. We assumed a thin disc composed of neutral hydrogen and helium, and multiplied the \hi\ surface densities by a factor of 1.4 to account for the helium and heavier elements. The total baryonic contribution to the observed rotation of the galaxy can therefore be written
\begin{equation}
v_{\rm baryonic} = \sqrt{\Upsilon_\star\,|v_{\rm disc}|v_{\rm disc} + |v_{\rm gas}|v_{\rm gas}}\,,
\end{equation}
where $\Upsilon_\star$ is the stellar mass-to-light ratio (M/L) of the disc, $v_{\rm disc}$ the disc velocity and $v_{\rm gas}$ the velocity of the gaseous disc.

From Jarrett et al. ({\it in prep.}), we have obtained a W1-W2 colour of $0.089\pm0.040$ mag; following the $M_\star/L_{\rm W1}$ calibration from \citet{Cluver2014}, we derived a stellar mass-to-light ratio of $0.62\pm0.11$ \ml. We will hereafter refer to this value as the {\it WISE} colour M/L value.

\subsubsection{Dark matter component}
The other major component in disc galaxies -- besides the baryonic components -- accounting for their total mass is the dark matter (DM). Indeed, several studies have shown that an important fraction of galaxies' mass -- especially for late-type galaxies -- is contained in their DM halo \citep[e.g.,][]{Rubin1978,Kalnajs1983,Carignan1985}. Two main profiles are generally used to describe the halo of this unseen matter: the pseudo-isothermal model which assumes an approximately constant DM density in the centre of the DM halo, and the {\it cusp-like} profile \citep{Navarro1996,Navarro1997}, commonly referred to as the Navarro-Frenk-White (NFW) profile, for which the DM density rapidly increases towards the centre of the halo. The implication of a constant DM density like in the case of the pseudo-isothermal profile is that it results in a rotation curve that increases linearly with radius, while the NFW profile gives rise to a square-root-like curve. Given the solid body-type rotation of IC 2574, we opt to use the pseudo-isothermal model to describe the galaxy's DM halo. The mass density and velocity distributions of the pseudo-isothermal sphere as a function of the radius $r$ of the sphere are respectively given by
\begin{eqnarray}
  \rho_{\rm iso}(r) &=& {\rho_0 \over 1+(r/r_{\rm c})^2},\\
  v_{\rm iso}(r) &=& \sqrt{4\pi G\rho_0 r_{\rm c}^2\left[1-(r/r_{\rm c})\arctan{(r/r_{\rm c})}\right]}
\end{eqnarray}
where $\rho_0$ and $r_{\rm c}$ are respectively the central density and the core radius of the halo. 

The decomposition of IC 2574's rotation curve was done using the \gipsy's program \rotmas. The weight affected to a given data point of the rotation curve is equal to the squared inverse of the error bar at that point.

When the stellar M/L is let as a free parameter (best fit model), the least-square fitting of the pseudo-isothermal model to the rotation curve returns a M/L value of $\Upsilon_\star = 0.20\pm0.10$ \ml, with a halo core radius and central density of $r_{\rm C} = 5.8\pm0.6$ kpc and $\rho = (6.7\pm1.0)\times10^{-3}\rm\, \Mo\,pc^{-3}$, respectively (see \Cref{tb:massmodel}). The obtained value of the M/L ratio is about 3 times lower than that inferred from the {\it WISE} colour of the galaxy, but is in agreement with the values obtained in \citet{Martinsson2013} and \citet{Lelli2016}. In fact, by measuring the vertical velocity dispersion of disc stars in a sample of 30 spiral galaxies, the authors determined a $3.6\rm\mu m$ stellar M/L ratio of about 0.2 \ml. At the last measured point (18.4 kpc), the dark component is 10.7 times more massive than the luminous (stars \& gas) component. In other words, the dark halo constitutes over 91\% of the total dynamical mass of IC 2574, meaning that the stellar and \hi\ discs of the galaxy contribute very little to its total mass. This result is consistent with the study by \citet{Martimbeau1994}, who found a dark to luminous mass ratio of 8.9 at their (distance-corrected) last radius of 10.8 kpc, and such large contribution of the dark halo was found in other low mass systems (e.g., DDO 154: \citealt{Carignan1988,Carignan1989}; NGC 3109: \citealt{Jobin1990,Carignan2013}).

To check the effect of the extension of the galaxy's rotation curve on the DM halo model, we have run the fit on the THINGS-only rotation curve; the halo parameters obtained from the fit were sensibly similar to those obtained with the extended curve, showing that the combination has little effect in constraining the model. Because this could be due to the large error bars of the low-resolution DRAO curve, we have used a uniform weighting scheme for the least-square fitting but obtained a similar result.
We performed an additional analysis by combining the DRAO curve to the \citet{Oh2008}'s IWM rotation curve of the galaxy instead of the bulk rotation curve (see \Cref{fig:adrift}). The rising part of the IWM rotation curve is slightly shallower than the bulk curve, and has a solid body type of rotation at almost all radii. Because of this shape, the best fit model with the pseudo-isothermal profile fails to constrain the parameters of the DM halo, yielding a large core radius of $24.4\pm4.5$ kpc with a stellar M/L ratio of $0.23\pm0.02$ \ml. Moreover, when the M/L ratio is fixed to the {\it WISE} inferred value, the core radius considerably increases to $4.9\e{7}$ kpc, which is unphysical. It is worth noting that the NFW model also fails to constrain the DM halo. When combined with the DRAO curve following the prescription in \Cref{sec:massmodel}, the curve becomes slightly flat in the outer regions, allowing us to constrain the DM halo with the pseudo-isothermal model. The best model fit provides a core radius of $9.7\pm1.0$ kpc with a M/L ratio of $0.13\pm0.06$ \ml, while the fixed M/L method provides a core radius of $16.6\pm3.4$ kpc. Again, we note that the {\it WISE} inferred M/L value overestimates the baryonic contribution to the total mass of the galaxy.

We then fixed the stellar M/L ratio to the {\it WISE} colour value, and fitted once again the pseudo-isothermal model to the rotation curve. As \Cref{tb:massmodel} shows, the values obtained for the halo parameters do not significantly vary, although a comparison of the reduced $\chi^2$ values suggests that the fixed M/L fit is less accurate than the best fit model. 

For completeness, we have also used the cusp-like Navarro-Frenk-White \citep[NFW;][]{Navarro1996,Navarro1997} density distribution to decompose the rotation curve of the galaxy. Similarly to \citet{Oh2008} and \citet{DeBlok2008}, we have obtained a value of zero, therefore unphysical, value for the stellar M/L. The same result was obtained by previous authors for similar galaxies (e.g., NGC 5585, \citealt{Cote1991}; NGC 3109, \citealt{Carignan2013}) and for a significant sample of galaxies \citep{Haghi2018}. This implies that the DM halo of the galaxy presents no cusp, and the pseudo-isothermal model should be preferred to the NFW model for the description of the galaxy's DM halo. To investigate whether the inclusion of stellar feedback in the NFW model can produce an acceptable fit to the rotation curve of the galaxy, we have used the criteria of \citet{DiCintio2014}. The authors found a dependence of the inner slope of the dark matter density profile, $\alpha$, on the stellar-to-halo mass ratio ($\rm\Ms/M_{halo}$). More specifically, for a galaxy like IC 2574, whose $\rm\Ms/M_{halo}$ value is less than ${\sim}0.01$ percent (\Cref{tb:massmodel} suggests a ratio of $0.001$ percent), the energy from stellar feedback is insufficient to significantly alter the inner dark matter density, and therefore the galaxy retains a cusp-like profile. Also, even for galaxies where stellar feedback is sufficient to produce cores in the centre of the halo, the core radius found is only a few kpc in size, definitely lower than the values obtained here. Based on these criteria, we conclude that the NFW profile is not suited to describe the DM profile of IC 2574.

\begin{figure*}
\makebox[\textwidth][c]{
\hspace{10pt}
\includegraphics[width=1.15\textwidth]{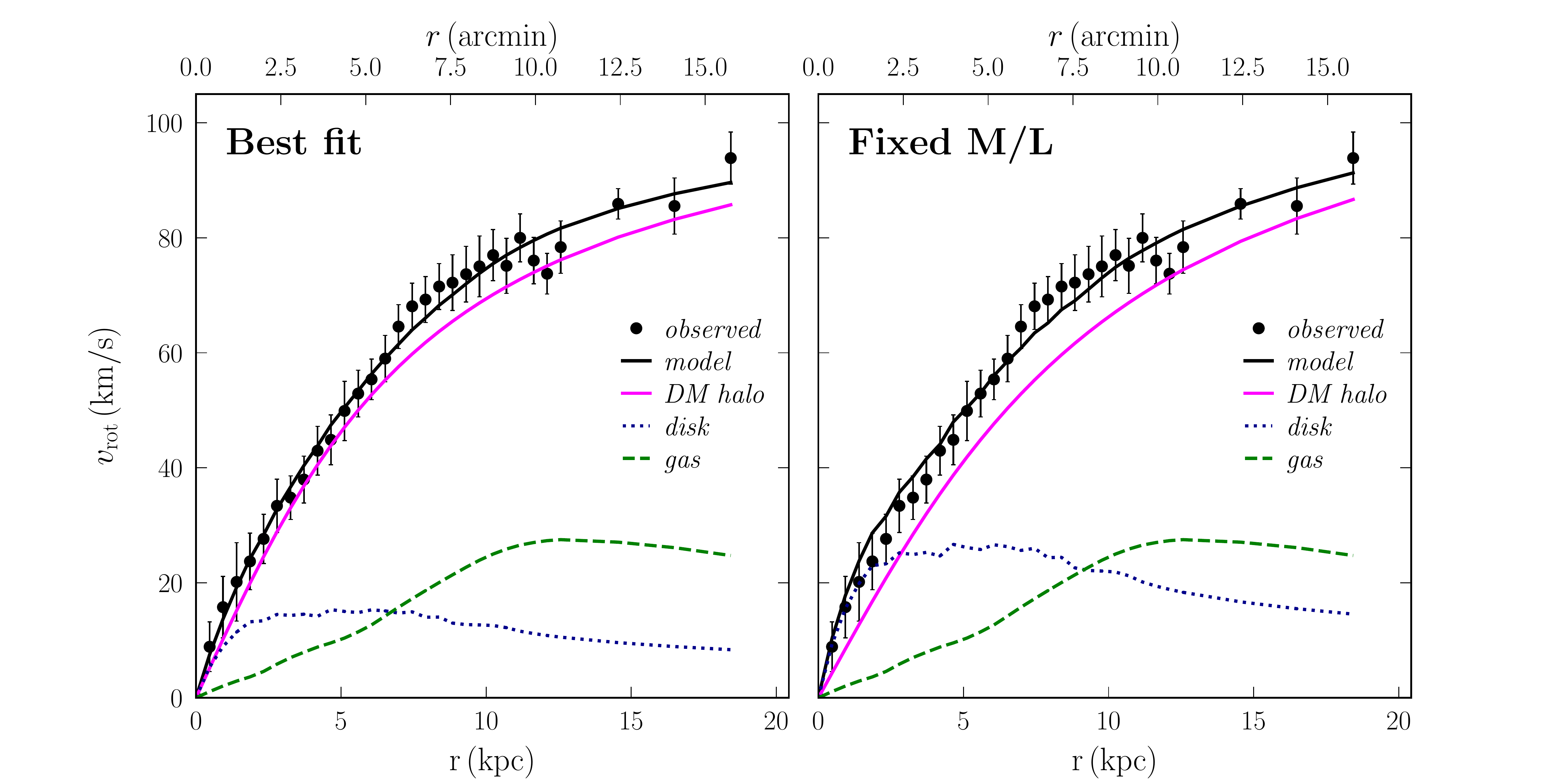}}
\vspace{-15pt}
\caption{Mass models of IC 2574 for the best fit ({\it left panel}) and fixed M/L ({\it right panel}) models of the pseudo-isothermal DM halo profile.}\label{fig:massmodel}
\end{figure*}

In \Cref{fig:massmodel} we show the different decompositions of the hybrid rotation curve with the pseudo-isothermal model.

\begin{table}
\centering
	\begin{tabular}{p{4.0cm} c c}
	\hline
    \hline
    Parameter & Best fit & Fixed M/L \\
    \hline
    \rule{0pt}{2.8ex}
    THINGS bulk + DRAO\\
    $\Upsilon_\star$ (\ml)\dotfill & $0.20\pm0.14$ & 0.62\\
    $r_{\rm c}$ (kpc)\dotfill & $5.79\pm0.65$ & $7.60\pm0.62$\\
	$\rho_0$ ($10^{-3}\,M_\odot\rm\,pc^{-3}$)\dotfill & $6.71\pm1.00$ & $4.66\pm0.37$\\
	$\chi^2_{\rm red}$\dotfill & 0.42 & 0.57 \rule[-1.7ex]{0pt}{0pt}\\
	At the last radius ($r=18.43$ kpc):\\
	$M_{\rm dark}/M_{\rm lum}$\dotfill & $10.7$ & $9.1$ \rule[-1.7ex]{0pt}{0pt}\\
    \hline
	\rule{0pt}{2.8ex}
    THINGS IWM + DRAO\\
    $\Upsilon_\star$ (\ml)\dotfill & $0.13\pm0.06$ & 0.62\\
    $r_{\rm c}$ (kpc)\dotfill & $9.71\pm0.95$ & $16.64\pm3.43$\\
	$\rho_0$ ($10^{-3}\,M_\odot\rm\,pc^{-3}$)\dotfill & $3.29\pm0.29$ & $2.11\pm0.20$\\
	$\chi^2_{\rm red}$\dotfill & 0.31 & 0.99 \rule[-1.7ex]{0pt}{0pt}\\
	At the last radius ($r=18.43$ kpc):\\
	$M_{\rm dark}/M_{\rm lum}$\dotfill & $11.1$ & $9.4$ \rule[-1.7ex]{0pt}{0pt}\\
    \hline   \rule{0pt}{2.8ex}
	$M_{\rm total}\,(M_\odot)$\dotfill & \multicolumn{2}{c}{$3.6\e{10}$} \rule[-1.7ex]{0pt}{0pt}\\
    \hline   
	\end{tabular}
	\caption{Results of IC 2574's mass models.}\label{tb:massmodel}
\end{table}

\subsection{\hi\ kinematics of HIJASS J1021+68}\label{sec:hijass_kinematics}
In \Cref{fig:hijass_velmaps} we show the velocity distribution in HIJASS J1021+68. There is a slight east-west rotation pattern in the complex, with the western side approaching the observer while the eastern side recedes. Also, the eastern clump seems kinematically separated from the rest of the complex, with a difference in velocity reaching ${\sim}10-20$ \kms. In the central and northern clouds of the complex, the velocity dispersion seems to follow the pattern of the velocity field, increasing from west to east. In the eastern cloud, the dispersion is roughly uniformly distributed with an average of ${\sim}20-25$ \kms. Although the limited resolution of the data does not allow us to perform an in-depth kinematical analysis of the complex, the present maps show that its kinematics are not exclusively supported by rotation nor is it solely supported by dispersion; instead, the data suggest that both the rotation and the random motions of the gas complex contribute equally to its gravitational support.

\subsection{The nature of HIJASS J1021+68}\label{sec:hijass_nature}
\begin{figure*}
\makebox[\textwidth][c]{
\includegraphics[width=0.9\textwidth]{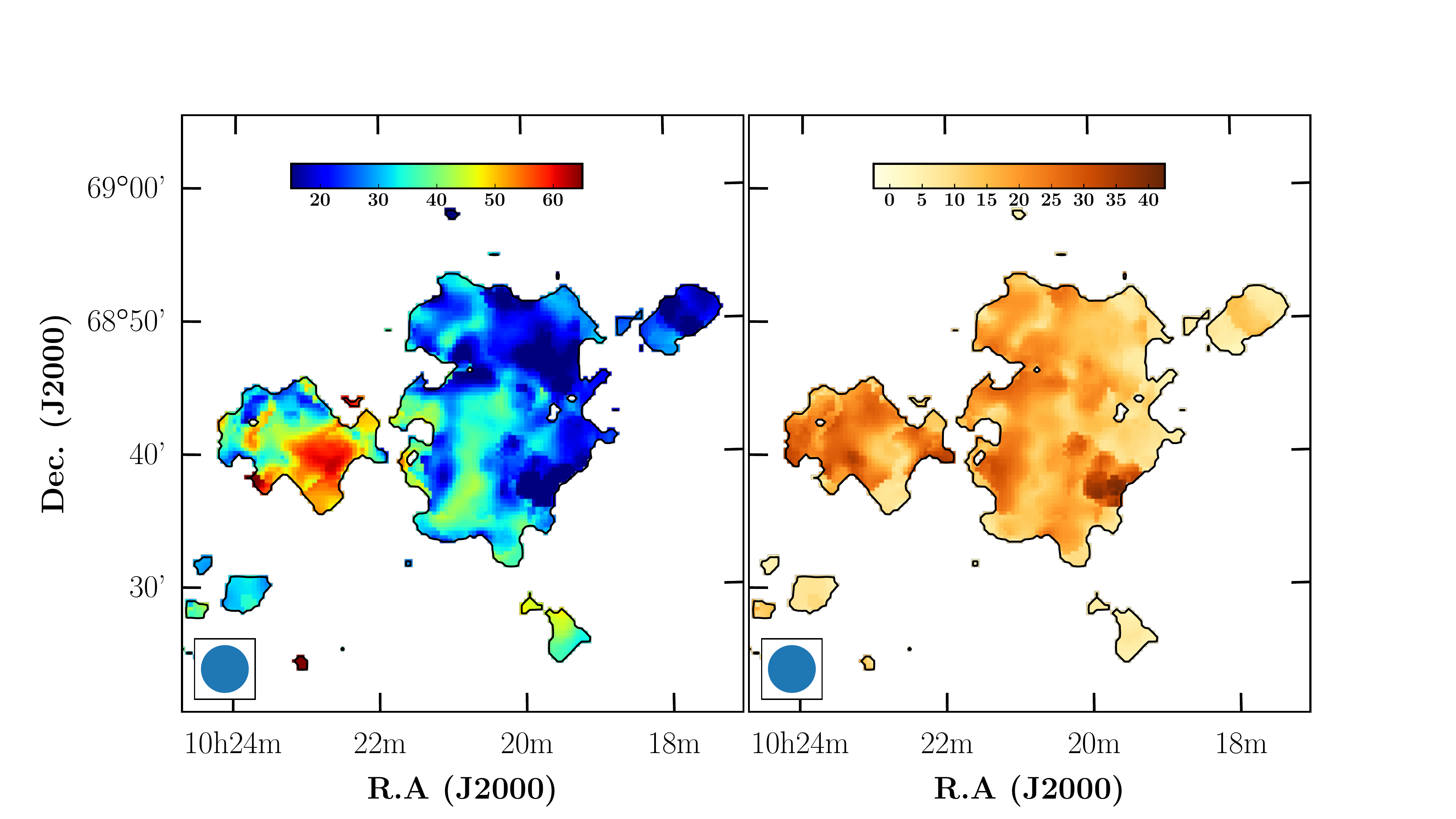}}
\vspace{-20pt}
\caption{Velocity field ({\it left panel}) and velocity dispersion map ({\it right panel}) of HIJASS J1021+68.}\label{fig:hijass_velmaps}
\end{figure*}

On the nature of HIJASS J1021+68, two possibilities arise: the object is either just a complex of gas clouds, or it is a dark galaxy in which star formation has not begun yet. The first hypothesis is obvious, as the system shows no stellar counterpart, and appears to be fragmented in different clouds. However, a few observational facts argue in favour of the dark galaxy nature of the object: the apparent gradient in the velocity field and the high velocity dispersion (up to ${\sim}20-25$ \kms) in the atomic gas, and the similarity of the object with the low-mass dwarf irregular galaxy GR8 \citep{Carignan1990}. In fact, the \hi\ kinematics of HIJASS J1021+68 present the same characteristics as those of GR8, whose gravitational support is essentially provided by rotation only in the inner regions, while it is dominated by random motions in the outer parts. Moreover, the rotation axis of the object -- as given by the gradient of the velocity in \Cref{fig:hijass_velmaps} -- roughly follows the north-south direction; the spatial distribution of the \hi\ in the system suggests that its major axis also follows the north-south direction. This makes HIJASS J1021+68 a system in which the rotation axis is parallel to the major axis, and the same pattern was observed in GR8 \citep[see][]{Carignan1990}. These kinematical similarities with GR8 show that HIJASS J1021+68 can possibly be a dark galaxy as first suggested by \citet{Boyce2001}.
However, for this possibility to be true, there has to be a connection between the eastern and central clouds, such that the whole complex forms a single body. Otherwise, the velocity gradient in the central cloud alone (of ${\sim}10$ \kms) is not high enough for the gravitational force to be important. Could there be some low column density gas in the space between the two major parts of the complex that was missed by the present observations?
To investigate the connections between the clouds in the complex and with the dwarf galaxy IC 2574, we plot in \Cref{fig:ic_hijass_nhi} a position-velocity (PV) diagram taken across the two systems. To maximise the signal-to-noise ratio, the width of the slice was taken to be $15'$, wide enough to cover most of the gas complex. Nonetheless, no apparent connection is seen between the major two parts of HIJASS J1021+68. However, the diagram reveals a set of small clouds (above the $2\sigma$ level) in the space between the two systems, most of which are not seen in the \hi\ intensity map of \Cref{fig:ic_hijass_nhi}. It is worth noting that the distribution of these clouds suggests that more non-Galactic clouds could exist in the contaminated velocity channels (grey band in \Cref{fig:ic_hijass_nhi}), that were removed by the Galactic \hi\ subtraction process. We note two elongated prominent structures in the contaminated region, respectively in IC 2574 and below HIJASS J1021+68 (at velocities of ${\sim}0$ \kms), which represent residuals of the Galactic \hi\ at the peak contamination.
Furthermore, the continuity of velocity in the distribution of the detected clouds suggests that the connection that the low resolution HIJASS data \citep{Boyce2001} hinted at may be real. The western region of IC 2574's extended envelope is also seen on the western side of the galaxy, at an angular offset of  about $-40'$.
Although HIJASS J1021+68 was previously classified as a dark galaxy and its velocity field presents a slight east-west gradient, the morphology of the complex and its apparent connection to IC 2574 strongly suggests that it may not be a dark galaxy. One possibility that could arise is that the object is, as suggested by \citet{Boyce2001}, a forming tidal dwarf galaxy in which star formation has not yet begun. However, as predicted by the simulations \citep[e.g.,][]{Bournaud2006}, tidal dwarf galaxies are more likely to form in major, gas-rich mergers. More specifically, tidal dwarf galaxies are known to generally form out of the debris residing in tidal tails of galaxies, as a result of interactions \citep[e.g.,][]{Duc2000,Duc2004,Lelli2015,Lee-Waddell2016}. This makes it less likely for HIJASS J1021+68 to be a tidal dwarf galaxy, since neither the optical nor the \hi\ morphology of IC 2574 hints at the existence of a tidal tail associated with the galaxy. With the natures of dark galaxy and tidal dwarf galaxy discarded as possible natures of HIJASS J1021+68, it becomes most likely that the object is just a complex of gas clouds connected to the large envelope of IC 2574, probably stripped from, or falling onto the envelope.

%


\section{Summary}\label{sec:summary}
In this work we have used sensitive DRAO observations of IC 2574 and HIJASS J1021+68 to map both the low and high column density gas in the region. The high sensitivity \hi\ intensity map of IC 2574 revealed a substantial amount of \hi\ around the galaxy, in the form of an \hi\ envelope, two large concentrations south and west of the galaxy, and as a series of small \hi\ clouds in its immediate surroundings. The analysis of the \hi\ in the envelope shows that the detected gas is real and distinct from Galactic \hi. Also, the velocity distribution of the \hi\ in the envelope and in the clumps suggests that the low column density gas rotates with the galaxy, although a gradient is observed in the velocity distribution.

Surprisingly, no clear \hi\ emission is detected in the space between M81 and HIJASS J1021+68, despite the apparent continuity of velocity between the two objects. This makes it unlikely that the system formed by IC 2574 and HIJASS J1021+68 is connected to the central members of the M81 group. However, if IC 2574 has interacted or connected to M81 at some point in the past, this could make it possible for the gas between the two systems to be ionised.

Using a 3D tilted-ring model, we derived the rotation curve of the galaxy to a larger extent than previous works. Taking advantage of higher resolution curves available in the literature, we constructed a hybrid rotation curve by combining our extended curve to that derived from the THINGS bulk velocity field in \citet{Oh2008}. This provided an extended and high resolution rotation curve with a slightly flat part in the outskirts, which can be used to constrain the DM halo parameters. We then used the pseudo-isothermal model to decompose the resulting rotation curve, and obtained a stellar M/L ratio of $0.2\pm0.1$ \ml. This is consistent with the morphological type of the galaxy and with studies in the literature \citep[e.g.,][]{Lelli2016}.

The observations also allowed to resolve the object HIJASS J1021+68 for the first time, and its \hi\ intensity map revealed three major clumps forming the complex. A search for a counterpart in UV, optical R-band and infrared {\it WISE} W1 was unsuccessful, and the \hi\ kinematics of the system suggests that both rotation and random motions contribute equally to its gravitational support. Furthermore, our analysis of a PV diagram of the data shows strong evidence of an \hi\ connection between the gas complex and IC 2574, but not between the central and eastern clouds of the complex. This sheds light on the nature of HIJASS J1021+68, and we argue that the object is not a dark galaxy as previously thought, but a complex of \hi\ clouds stripped from, or falling onto the envelope of IC 2574.

\section*{Acknowledgments}
The authors are very grateful to the staff at DRAO, particularly to Operations Manager Dr. Andrew Gray for his flexibility and assistance in observing the many fields and gathering archived data for our work.
We would like to thank the anonymous referee, whose comments and suggestions were very useful and helped to improve the quality of the present work.
The work of CC is based upon research supported by the South African Research Chairs Initiative (SARChI) of the Department of Science and Technology (DST), the South African Radio Astronomy Observatory (SARAO) and the National Research Foundation of South Africa (NRF). The research of AS has been supported by SARChI and SARAO fellowships. The research of LC is supported by the Comit\'{e} Mixto ESO-Chile and the DGI at University of Antofagasta.

\bibliographystyle{mnras}
\bibliography{library}

\bsp	
\label{lastpage}
\end{document}